\documentclass[journal]{IEEEtran}

\pdfoutput=1
\usepackage{ifpdf}

\usepackage[compress]{cite}

\ifCLASSINFOpdf
  \usepackage[pdftex]{graphicx}
  \DeclareGraphicsExtensions{.pdf,.jpg}
\else
  \usepackage[dvips]{graphicx}
  \DeclareGraphicsExtensions{.eps}
\fi

\ifCLASSOPTIONcompsoc
  \usepackage[caption=false,font=footnotesize,labelfont=sf,textfont=sf]{subfig}
\else
  \usepackage[caption=false,font=footnotesize]{subfig}
\fi

\usepackage[cmex10]{amsmath,mathtools}
\usepackage{amssymb}
\usepackage{mathrsfs}
\usepackage{dsfont}
\usepackage{mathdots}

\usepackage{algorithmic}

\usepackage{array}
\usepackage{blkarray}

\usepackage{stfloats}

\usepackage{url}

\usepackage[usenames,dvipsnames,svgnames,table]{xcolor}
\usepackage{pifont}
\usepackage{fancybox}
\usepackage{todonotes}
\usepackage{soul}
\usepackage{hhline}
\usepackage{multirow}
\usepackage{extarrows}
\usepackage{epstopdf}
\usepackage{tabu}
\usepackage{multirow}

\begin{document}

\title{Analyzing the Synergy between Broadcast Polling and Piggybacking in WiMAX Networks}
\author{Maria Iloridou,~Evangelos~Papapetrou,~\IEEEmembership{Member,~IEEE}, and~Fotini-Niovi~Pavlidou~\IEEEmembership{Senior Member,~IEEE}
\thanks{M. Iloridou and F.-N.Pavlidou are with the Department of Electrical \& Computer Engineering, Aristotle University of Thessaloniki, Thessaloniki, 54636, Greece (e-mail:iloridou@auth.gr,niovi@auth.gr)}
\thanks{E.Papapetrou is with the Department of Computer Science \& Engineering, University of Ioannina, Ioannina, 45110, Greece (e-mail:epap@cse.uoi.gr)}
}

\ifCLASSOPTIONpeerreview
\IEEEpeerreviewmaketitle
\else
\maketitle
\fi

\begin{abstract}
In this work we present a model for analyzing the combined use of broadcast polling and piggybacking in Worldwide Interoperability for Microwave Access (WiMAX) networks. For an accurate analysis of piggybacking, the model focuses on the realistic case of limited up-link bandwidth and non-trivial queueing capability at the subscriber stations. We first model the activity of a subscriber station using a Markov chain and its queue as an M/G/1 system with vacations in order to facilitate the analysis of the piggyback mechanism. We then derive a set of fixed point equations that describe not only the contention process at the network level but also bandwidth allocation to contending and piggybacked requests. Our model uses a minimal set of assumptions and is generic in the sense that it is customizable through a set of parameters. It can also reproduce the system performance in both saturated and non-saturated conditions. After validating our analysis through extensive simulations, we shed light on the aspects of the synergy between broadcast polling and piggybacking and unveil the pros and cons of using the latter.  
\end{abstract}

\ifCLASSOPTIONpeerreview
\else
\begin{IEEEkeywords}
piggyback, broadcast polling, IEEE 802.16
\end{IEEEkeywords}
\fi

\section{Introduction}

\IEEEPARstart{W}{orldwide} Interoperability for Microwave Access (WiMAX) is probably the most promising broadband wireless technology for the next decades. WiMAX has been standardized by the IEEE 802.16 Working Group~\cite{ieee2009ieee}. To meet the service requirements of various applications the \emph{medium access control (MAC)} sublayer of the IEEE 802.16 standard supports five scheduling services in the popular \emph{Point-to-Multipoint (PMP)} mode of operation: the unsolicited grant service (UGS), the real-time Polling Service (rtPS), the extended real-time Polling Service (ertPS), the non-real-time Polling Service (nrtPS) and the Best Effort (BE) one. Among these services BE holds a central role not only due to its inherent compatibility with the Internet architecture but also because it is the choice for a variety of WiMAX application classes such as media content downloading, web browsing, instant messaging, etc~\cite{wimax-survey}. According to the standard, the mandatory medium access method for BE follows a contention-based approach that features a bandwidth request (BR)-grant scheme. In other words, subscriber stations (SSs) with BE data to transmit should contend for receiving a bandwidth (BW) grant from the base station (BS). 

In this work we focus on the realization of the contention-based access scheme for BE in the widespread scenario of fixed access networks, also known as \textit{broadcast polling}~\cite{ieee2009ieee}. Analyzing this mechanism is of paramount importance not only due to the key role of BE but also because it is used by other scheduling services such as nrtPS and ertPS~\cite{ieee2009ieee,wimax-survey}. Reasonably, several researchers looked into analyzing broadcast polling~\cite{vinel2005performance,vinel2006wlc22,modeling_hsiao,fallah-model,bandwidth_request_hsiao,desing_analysis_tseng,vu2010performance,chuck2010comprehensive,liu2012performance,giambene2013nonsaturated} (as well as its variant, group/multicast polling~\cite{unsaturated_ni,unsaturated_ni_systems}) and provided valuable insight into its characteristics. Yet, SSs implementing broadcast polling may also use on top of it the optional \textit{piggyback} mechanism.  Albeit the latter is a cost-effective alternative that can provide significant performance improvements~\cite{bandwidth_request_hsiao,cicconetti2007performance}, so far its analysis has received little attention~\cite{seo-steady-globecom06,lee2007queueing,desing_analysis_tseng,bandwidth_request_hsiao}. The proposed models are a first step towards evaluating piggybacking but unfortunately cannot accurately capture all of its performance features.
There are two essential reasons for this, besides the ones relating to the modeling of the underlying contention mechanism (we discuss this latter issue in Section~\ref{related}). 
First, the proposed models assume either no queueing capability~\cite{bandwidth_request_hsiao} or a trivial one (just data from a single arrival)~\cite{desing_analysis_tseng}. Consequently, it is not possible to model piggybacking when it is mostly needed, i.e. when an SS's queue builds up due to data arrival bursts. The second major drawback is that unlimited up-link (UL) BW is assumed~\cite{seo-steady-globecom06,lee2007queueing,desing_analysis_tseng}. This makes it impossible to model the coupling between piggybacking and the contention mechanism that results from sharing UL BW. 

Motivated by these observations we focus on analyzing broadcast polling with piggybacking in a setting with limited UL BW, i.e. we wish to model not only the transmission of BRs but also the BW allocation phase. To this end, we follow a typical approach used in the analysis of broadcast polling~\cite{fallah-model,giambene2013nonsaturated}, i.e to model an SS's activity with a Markov chain. Then, we capitalize on concepts from Bianchi's seminal work~\cite{bianchi} on IEEE 802.11 networks. Nonetheless, we generalize the Markov chain so as to include a branch that models the piggyback process. More importantly, to facilitate an accurate analysis of piggybacking we consider SSs with non-trivial queueing capability which we model using an M/G/1 system with vacations.
In summary our contributions are:
\vspace{-4pt}
\begin{itemize}
	\item We propose a customizable analytical model for broadcast polling with piggybacking (Sections~\ref{SS-activity-markov} and~\ref{network-analysis}). The analysis covers both BR transmission and BW allocation phases. It also features a set of realistic characteristics that are critical for modeling piggybacking such as limited UL BW and SSs with non-trivial queueing capability.
	\item Our analysis is also suitable for investigating plain broadcast polling, i.e. without piggybacking. Compared to other approaches in the literature, our model provides a more detailed modeling of the BW management mechanism (e.g. time window for serving a BR, number of SSs served per frame, etc).
	\item We provide a detailed performance evaluation of piggybacking using both simulation and analytical results (Section~\ref{evaluation}).
	\item We confirm that piggybacking can potentially bring performance improvements but we find that this depends on the ratio of BW allocated for contention to that allocated for data transmission. We discuss the optimal strategy and at the same time we shed light on the trade-offs involved in using piggybacking (Section~\ref{evaluation}).
\end{itemize}

\noindent As a final note, we believe that our analysis could be a valuable tool when examining piggybacking over the contention-based access scheme for mobile access WiMAX networks since the latter shares striking similarities with broadcast polling.

In the rest of the paper, we provide a brief overview of the contention-based access mechanism for fixed WiMAX networks (Section~\ref{overview})  and review the related literature (Section~\ref{related}). In Section~\ref{assumptions}, we present the system model and the assumptions considered in our analysis. We conclude this work in Section~\ref{conclusions}.

\section{Related Work}
\subsection{Overview of Contention-based Access with Piggyback}\label{overview}

As mentioned previously, in PMP WiMAX networks, the mandatory access scheme for SSs with BE connections uses a BR-grant schema in a contention-based mode. In other words, an SS, wishing to receive a BW grant for sending data, should first contend for sending to the BS a BR with its BW needs. Regarding the contention process, there are two realizations; one proposed for SC and OFDM physical layer (PHY) specifications, i.e. for fixed access networks, and another one for use with OFDMA PHY utilized for mobile access. The two implementations share many similarities. In the OFDMA case, SSs contend through the transmission of a code in the ranging region of a frame (\textit{Contention-based CDMA BRs}). On the other hand, in fixed access networks all SSs (\textit{broadcast polling}) or a group of them (\textit{group polling}) contend by transmitting BRs in a period of time allocated by the BS in each UL subframe. 

In the case of fixed access networks, the contention period is organized in transmission opportunities (TOs), the realization of which depends on the PHY specification. An SS with data to send should wait for a random number of TOs before sending the BR to the next one. This number is uniformly selected from the interval [0,$W_{0}$-1], where $W_{0}$ is known as the initial contention window size. Note that the waiting period may span multiple frames depending on  $W_{0}$. If two or more SSs choose the same TO to transmit a BR then a collision occurs. Upon correctly receiving a BR, i.e. no collision occurs, the BS should inform the SS about a BW grant through the UL-MAP in the DL subframe. Since BW is not always available, the BS may provide a grant not necessarily in the next frame but within a number of frames following the one that the successful BR transmission took place. This number is known as the Contention-based Reservation Timeout or the T16 period. The standard does not specify neither carrier sensing nor an acknowledgment mechanism for SSs whose BRs have successfully been transmitted. Therefore, in the case of BR collision the SS will have to wait until the T16 period expires. Then it assumes a collision and retransmits its BR. 

To deal  with retransmissions the SS uses the \textit{truncated binary exponential back-off (TBEB)} algorithm. According to it, in the event of failing to receive a grant the SS doubles its contention window and retransmits the BR after deferring for a number of TOs. This number is randomly chosen from the interval [0, $W_{1}$-1], where $W_{1}\!=\!2W_{0}$ denotes the new window size. After the $i$-th failure the window size is $W_{i}\!=\!2^{\min\{m,i\}}W_{0}$. Here $m$ implicitly defines the maximum contention window  $2^{m}W_{0}$, which is specified in the Uplink Channel Descriptor (UCD) along with $W_{0}$. For each BR there is a maximum allowable number of retransmission attempts. If this limit is reached the data associated with the BR shall be dropped.

Finally, the standard also defines an optional \textit{piggyback} procedure. According to it, an SS, after receiving a grant, may request more BW by piggybacking a collision-free BR to the data instead of contending.

\subsection{Analytical models for Contention-based Access in WiMAX}\label{related}

Several researchers have looked into the contention-based access mechanism of IEEE 802.16 networks. Naturally, the proposed analytical models can be classified into two broad categories; one focuses on broadcast/group polling~\cite{vinel2005performance,vinel2006wlc22,modeling_hsiao,fallah-model,bandwidth_request_hsiao,desing_analysis_tseng,vu2010performance,chuck2010comprehensive,liu2012performance,giambene2013nonsaturated,unsaturated_ni,unsaturated_ni_systems} while the other studies the contention-based CDMA BRs mechanism~\cite{choi2006multichannel,staehle2009performance,seo2011design,seo2011queuing,liu2014performance,bianchi2014modeling}. Interestingly, most efforts in both categories do not consider the optional piggyback process even though it can be used with any of the aforementioned contention methods.

\begin{table*}[t]
	\caption{Network Parameters and Notation used in the Analysis}
	\label{analnottable}
	{\setlength\tabcolsep{2pt}
		\centering
		\begin{tabu}{|[0.8pt]p{0.05\textwidth}|p{0.43\textwidth}||p{0.05\textwidth}|p{0.43\textwidth}|[0.8pt]} 
			\tabucline[0.8pt]{-}
			\multicolumn{4}{|[0.8pt]c|[0.8pt]}{Network Parameters}\\
			\tabucline[0.8pt]{-}
			$N$ & number of SSs in the network & $L$ & number of data slots in the UL subframe \\
			\hline	
			$N_s$ & number of TOs in the UL subframe &  $W_{0}$ & initial contention window size ($\geq N_{s}$)\\
			\hline
			$2^{m}W_{0}$ & maximum contention window size & $W_i$ & contention window size in round $i$\\
			\hline
			$\mathrm{M}$ & maximum of frames in T16 period & $\mathrm{D}$ & maximum number of retransmissions for a BR\\
			\hline
			$\mathrm{G}$ & maximum number of piggybacked BRs & $\lambda$ & mean rate of packet generation at each SS\\
			\hline
			$\mathrm{T_{fr}}$ & frame duration & $\rho_{in}$ & ($=\lambda\mathrm{T_{fr}}$) offered load\\		
			\tabucline[0.8pt]{-}
			\multicolumn{4}{|[0.8pt]c|[0.8pt]}{Performance related notation}\\
			\tabucline[0.8pt]{-}
			$p$ & BR collision probability in a TO & $q$ & probability of receiving a BW grant in a frame\\
			\hline
			$q_{_{\mathrm{M}}}$ & probability of granting BW within $\mathrm{M}$ successive frames & $p_f$ & probability of not receiving BW in a contention round\\
			\hline
			$\mathrm{E}\{S\}$ & expected service delay & $\mathrm{E}\{S_{C}\}$ & expected service delay through contention\\
			\hline
			$\mathrm{E}\{S_{Q}\}$ & expected service delay seen by queued packets & $\mathrm{E}\{W\}$ & expected waiting delay\\
			\hline
			$\rho$ & utilization ($=\lambda \mathrm{E}\{S\}$) & $\bar{G}$ & mean number of packets transmitted with piggybacking\\
			\hline
			$\Pi_{0}$ & probability that an arriving packet finds an empty queue & $P_{\mathrm{B}}$ & probability that an SS transmits a BR in a frame\\
			\hline	
			$P_{\mathrm{S}}$ & probability that a TO contains a successful BR & $P_{\mathrm{D}}$ & probability of dropping a BR\\
			\hline						
			$Th$ & total throughput seen by an SS & $Th_{\mathrm{C}}$ & throughput achieved through the contention mechanism\\
			\hline						
			$Th_{\mathrm{G}}$ & throughput achieved through piggybacking &  & \\
			\tabucline[0.8pt]{-}
		\end{tabu}
	}
	\vspace{-12pt}
\end{table*}
Overall, the analysis of piggybacking has received little attention~\cite{seo-steady-globecom06,desing_analysis_tseng,lee2007queueing,bandwidth_request_hsiao} despite the fact that it can bring significant performance improvements~\cite{bandwidth_request_hsiao,cicconetti2007performance}. More specifically, the first modeling attempt assumes the contention-based CDMA BRs as the mandatory mechanism~\cite{seo-steady-globecom06,lee2007queueing}. The authors consider the queueing performance of an SS with an exhaustive queue service. To this end, they use an M/G/1 model and a Markov chain for modeling transmission periods. The authors also consider two disciplines for bandwidth allocation; transmitting one~\cite{seo-steady-globecom06} or multiple~\cite{lee2007queueing} packets per BW grant. Although the proposed model brought some important concepts to the analysis of piggyback, some of its assumptions significantly limit its accuracy. The most important is the assumption that the available UL BW is unlimited. In a real system the mandatory contention-based mechanism shares the same limited BW with piggybacking. Therefore the operation of one interferes with the other and vice-versa. For example, the amount of BW allocated for piggybacking influences the probability that the BS will grant BW to a contention-based BR. This in turn affects the probability that the requesting SS will issue another BR. Thus, assuming unlimited UL BW results in a less realistic modeling because it is not possible to capture the coupling between piggybacking and the contention mechanism. Moreover, the authors determine the piggyback probability based on the average queue size instead of the actual queue status. They also consider an exhaustive service, i.e. piggyback is utilized until the SS's queue is empty. This assumption deviates from a real system especially if we consider that under limited BW such a policy allows SSs using piggyback to drive contending ones to starvation.
Another effort for analyzing piggybacking, this time using broadcast polling as the mandatory mechanism, has been presented in~\cite{desing_analysis_tseng}. Still, this method also assumes unlimited UL BW and some sort of exhaustive service, i.e. there is no apparent limitation on how many times an SS may use piggybacking after receiving a BW grant. More importantly, the proposed model is actually semi-analytical since it requires that the request collision probability, a key performance index of broadcast polling, is determined by simulation. Finally, the authors implicitly assume a trivial queueing capability for each SS, i.e. the model considers only one of all possible arrivals during an SS's busy period. So far, the most realistic model has been proposed by He et.al.~\cite{bandwidth_request_hsiao}. It includes an analysis of broadcast polling, although somehow simplified (e.g. the T16 period is only one frame long). It also assumes limited UL BW as well as a more practical piggyback policy. Yet, the model bears a significant limitation; it assumes no queuing capability at the SSs and considers piggybacking only for transmitting packets that require BW allocation in successive frames. This is of paramount importance since piggybacking is mostly useful for transmitting queued packets.      

In this work we put emphasis on a more realistic modeling of piggybacking in order to unveil all aspects of its performance and its synergy with the mandatory broadcast polling mechanism under both saturated and non-saturated traffic. To this end, we assume limited UL BW as well as SSs with non-trivial queueing capability. This clearly differentiates our work from other efforts studying piggybacking. Moreover, our approach also provides an improved analysis of the broadcast polling mechanism. Indeed, so far, various researchers have provided analytical models that capture the performance of broadcast/group polling only in saturated conditions~\cite{vinel2005performance,modeling_hsiao,vu2010performance} or in unsaturated conditions with the assumption of unlimited BW~\cite{vinel2006wlc22,desing_analysis_tseng,unsaturated_ni,unsaturated_ni_systems}, i.e. they only consider the contention process and not the data transmission phase. Only a set of more recent studies assumes both saturated and non-saturated traffic conditions as well as limited UL BW~\cite{fallah-model,chuck2010comprehensive,giambene2013nonsaturated,bandwidth_request_hsiao,liu2012performance}. In this case one critical task is to accurately model the data transmission phase and predict the probability that the BS will not allocate a grant to a successful BR due to BW depletion. The models proposed in \cite{fallah-model} and \cite{chuck2010comprehensive} assume that the latter probability is constant and known beforehand. In \cite{giambene2013nonsaturated}, the probability is calculated only under the assumption that a single BR is served in each frame. Finally, in both \cite{bandwidth_request_hsiao} and \cite{liu2012performance}, a T16 period of one frame is assumed, i.e. only BRs received in one frame are considered for BW allocation and older BRs are dropped. This significantly affects the probability of receiving a grant. Our model moves one step forward by considering the most generic case, i.e. the BS provides grants to multiple BRs (both contending and piggybacked) in each frame while the T16 period can be greater than one frame.

\section{System model and Assumptions}\label{assumptions}
Our aim is to present a unified analytical model able to seamlessly portray the performance of broadcast polling with and without piggybacking. 
For this reason, we examine an IEEE 802.16 network operating in the PMP mode under either the SC or the OFDM PHY. We consider the scenario of a single BS and $N$ SSs. The frame is structured either in the FDD or the TDD mode and its duration is $\mathrm{T_{fr}}$. The UL subframe consists of $N_{s}$ TOs that SSs can use to transmit their BRs. For resolving collisions, the SSs implement the TBEB algorithm with initial window $W_{0}$, maximum window $2^{m}W_{0}$ and a maximum of $\mathrm{D}$ retries after which the BR is dropped. Since the optimal $W_{0}$ is not specified by the standard we reasonably assume that $W_{0}\!\geq\! N_{s}$. Otherwise the SSs will never choose the $N_{s}\!-\!W_{0}$ remaining TOs in their first transmission attempt. An SS can piggyback a BR on data for receiving a BW grant in the next frame. However, we assume that the SS can use piggybacking in up to $\mathrm{G}$ consecutive frames. This is a reasonable strategy, also used in the literature~\cite{bandwidth_request_hsiao}, in order to avoid driving other SSs to starvation. The UL subframe contains $L$ data bursts (hereafter called data slots for simplicity).
In general, $L$ is not sufficient for serving all successful BRs. Therefore the probability $q$ of providing a grant to a BR during a specific frame is not always one. Finally, we consider a T16 period of $\mathrm{M}$ frames.
Table~\ref{analnottable} summarizes the system parameters and the notation used in the analysis. Furthermore, we adopt the following assumptions:\\
\textit{(A1): Each SS has a single BE data connection.}\\
According to the IEEE 802.16 standard, BW is always requested on a connection basis but it is allocated to the SS. A1 allows us to focus on the access mechanism without going into the details that each manufacturer may implement for internal BW allocation at the SS. This assumption or a similar one, where multiple BE connections are treated as a single one, are common in the literature~\cite{vinel2006wlc22,unsaturated_ni,unsaturated_ni_systems,giambene2013nonsaturated,desing_analysis_tseng,fallah-model}.\\ 
\textit{(A2): Packets are generated according to a Poisson process (with rate $\lambda$) and buffered to the SS's queue. A BR is generated when a packet arrives at an empty queue or when the SS's queue is not empty and an opportunity for BR transmission exists, either through contention or piggybacking.}\\
We follow this reasonable approach since there is no universal algorithm for creating BRs based on the incoming traffic. An alternative would be to directly model the BR arrival process. The reasonable approach in this case is to assume a memoryless process~\cite{fallah-model,unsaturated_ni,unsaturated_ni_systems,giambene2013nonsaturated} unless a specific algorithm for producing BRs is known. However, note that an algorithm for producing BRs would certainly consider the system state, which results in an non memoryless process. Therefore, similar to many researchers~\cite{bandwidth_request_hsiao,desing_analysis_tseng,liu2012performance,vinel2009capacity}, we choose to directly model the packet arrival process.\\
\textit{(A3): An SS's queue is infinite.}\\
We make the assumption of an infinite queue, which is common in the literature~\cite{unsaturated_ni,unsaturated_ni_systems,giambene2013nonsaturated,modeling_hsiao,desing_analysis_tseng}, in order to explore the potential of piggybacking at its full extent. Note that in the reasonable case that the queue size is much greater than $\mathrm{G}$ then the limiting factor is actually $\mathrm{G}$ and the impact of the queue size is minimal.\\
\textit{(A4): The probability $p$ of a BR collision in a TO is constant and independent of the SS's retransmission history.}\\
This assumption is frequently found in the literature~\cite{bandwidth_request_hsiao,fallah-model,chuck2010comprehensive,unsaturated_ni,unsaturated_ni_systems,giambene2013nonsaturated,modeling_hsiao,vu2010performance,liu2012performance} and although it may not be entirely valid we will show, when validating our analysis, that its impact on the model accuracy is minimal.\\
\textit{(A5): The BS allocates BW grants of one data slot that correspond to the transmission of one data packet.}\\
Currently, there is no widely accepted bandwidth allocation mechanism for implementation at the BS. We adopt this frequently used assumption~\cite{modeling_hsiao,vu2010performance,liu2012performance,giambene2013nonsaturated,vinel2009capacity} in order to focus on the access mechanism without the need to examine fairness issues related to the bandwidth allocation problem. We feel that such issues are outside the scope of this work. Furthermore, note that for determining $q$, i.e. the probability of granting BW to a successful BR in a specific frame, the important parameter is the number of BRs that can be served in a frame. Including this assumption in our system model results in the BS being able to serve up to $L$ BRs per frame.\\
\textit{(A6): Pending BRs are served by the BS using a random order service discipline.}\\
The service discipline at the BS is important for determining $q$ when the BS can queue a BR and serve it within $\mathrm{M}\!\!>\!\!1$ frames. As mentioned, most research efforts assume either unlimited UL BW and thus $q\!\!=\!\!1$~\cite{vinel2005performance,vinel2006wlc22,desing_analysis_tseng,unsaturated_ni,unsaturated_ni_systems}, or a predefined $q$~\cite{fallah-model,chuck2010comprehensive}, or even limited UL BW with $\mathrm{M}\!\!=\!\!1$~\cite{vu2010performance,liu2012performance}. In all these cases the service policy is trivial. Only Giambene et. al.~\cite{giambene2013nonsaturated} assume a Round-Robin discipline for serving BRs from different SSs. In this case the service discipline is critical because they assume that only one SS may be served during a frame. Since we allow the BS to serve multiple SSs in each frame and each BR to be served within $\mathrm{M}$ frames, we expect the impact of a priority based discipline to be rather limited in our case. Therefore, we adopt the simple approach of randomly choosing the BR to serve which results in a constant $q$ that does not depend on the BR's waiting history. In general deciding on the optimal queue service discipline is a complex procedure that depends on the bandwidth allocation algorithm implemented at the BS. We believe that investigating this issue is outside the scope of this paper and plan to explore this in future work. 

In the following, our analytical approach is to first examine the medium access mechanism at the SS level, i.e. analyze the process implemented by an SS (Section~\ref{SS-activity-markov}). To this end, we model the SS as an M/G/1 queueing system with vacations and the medium access procedure, including piggyback, as a Markov chain. Later, in Section~\ref{network-analysis}, we combine the  Markov chain and well-established concepts from the analysis of IEEE 802.11 networks~\cite{bianchi} to investigate the performance of the random access scheme at the network level.
\begin{figure*}
	\centering
	\includegraphics[width=0.91\linewidth]{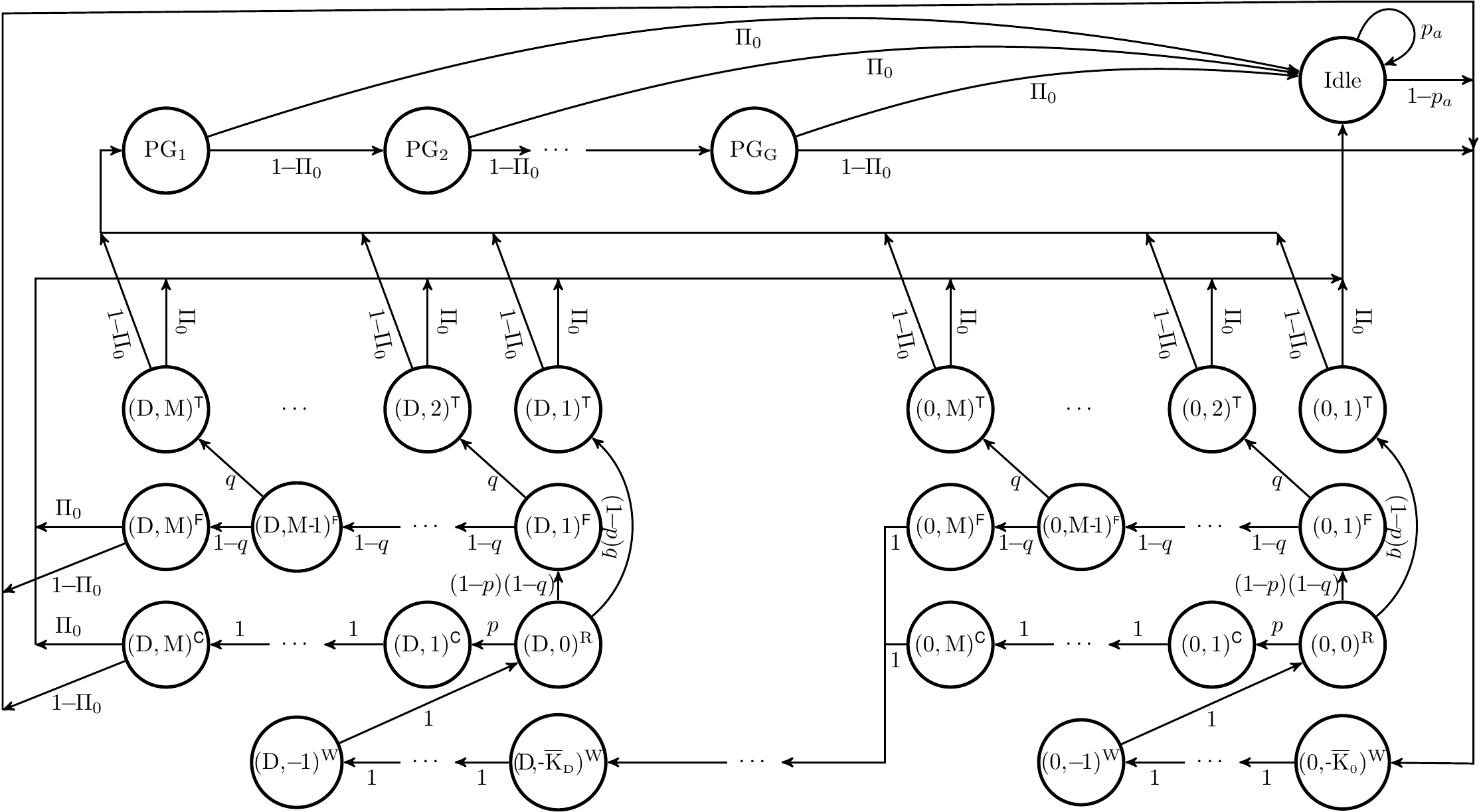}\\
	\caption{The proposed Markov chain}\label{markov_chain}
	\vspace{-12pt}
\end{figure*} 
\section{Modeling the activity of an SS}\label{SS-activity-markov}

To design the Markov chain that portrays the activity of an SS we first observe that each SS can be in one of the following phases during a specific frame:
\begin{description}
  \item[$\mathrm{W}$] waiting due to a back-off period
  \item[$\mathrm{R}$] transmitting a BR
  \item[$\mathrm{C}$] waiting due to a BR collision in a previous frame
  \item[$\mathrm{F}$] waiting for a BW grant without success after a successful BR in a previous frame 
  \item[$\mathrm{T}$] transmitting data after receiving a BW grant as a result of a successful BR in a previous frame
\end{description}
Let $\mathcal{S}=\{\mathrm{W},\mathrm{R},\mathrm{C},\mathrm{F},\mathrm{T}\}$ denote the set of the aforementioned phases.
Observe that each SS may contend in several rounds and each contention round consists of several frames. Furthermore, during a frame in a specific round an SS may be in any of the phases in $\mathcal{S}$. Therefore, in the proposed Markov chain we use a generic state \emph{$(i,j)^{\mathrm{X}}, X \in \mathcal{S}$} such that:
\newtheorem{definition}{Definition}
\begin{definition}[State $(i,j)^{\mathrm{X}}$]
	The state in which the SS is in phase $X$ during the $j$-th frame of the $i$-th contention round.
\end{definition}
For example, the interpretation of state $(i,j)^{\mathrm{F}}$ is that the SS did not receive any BW grant in the $j$-th frame of the $i$-th contention round although in a previous frame of the same contention round the SS transmitted successfully a BR. Furthermore, to model the piggyback mechanism as well as the case of an idle SS we also use the following states in the Markov chain:
\begin{description}
  \item[$\mathrm{PG}_{i}$] the SS transmits after receiving the $i$-th consecutive BW grant using the piggyback mechanism
  \item[$\mathrm{Idle}$] the SS waits until a new BR is produced
\end{description}
The proposed Markov chain is illustrated in Fig.~\ref{markov_chain}. When a BR is created, the terminal chooses a value for the back-off counter in the range $[0,W_{0}-1]$. Therefore, the SS may wait for more than a frame before the counter expires. Let $\overline{\mathrm{K}}_{0}$ denote the average number of the waiting frames before the frame in which the counter expires. We model the waiting period with the sequence of states $(0,j)^{\mathrm{W}}, j=-\overline{\mathrm{K}}_{0},\ldots, -1$ where the transition probability for traversing those states is one. 
After visiting the aforementioned states, the SS finally reaches state $(0,0)^{\mathrm{R}}$, which corresponds to the frame in which the back-off counter expires and the SS transmits the BR. This BR will be caught up in a collision with probability $p$, in which case the SS will wait for $\mathrm{M}$ frames (states $(0,j)^{\mathrm{C}}, j=1,\ldots,\mathrm{M}$ with transition probability from state $(0,j-1)^{\mathrm{C}}$ to state $(0,j)^{\mathrm{C}}$ being one) before moving to the next contention round. In the case that the BR is successful (probability $1-p$) the SS will move from state $(0,0)^{\mathrm{R}}$ either to state $(0,1)^{\mathrm{T}}$ or $(0,1)^{\mathrm{F}}$, depending on whether BW is granted (probability $q$) or not (probability $1-q$). 
Therefore, the transition probabilities from $(0,0)^{\mathrm{R}}$ to  $(0,1)^{\mathrm{T}}$ and $(0,1)^{\mathrm{F}}$ are $(1-p)q$ and $(1-p)(1-q)$ respectively. If no BW is granted (state  $(0,1)^{\mathrm{F}}$) the SS will wait for a BW grant in the next frame. Consequently, the SS will move to state  $(0,2)^{\mathrm{T}}$ or state  $(0,2)^{\mathrm{F}}$, with probabilities $q$ and $1-q$ respectively, depending on whether BW is granted or not.
The SS may wait up to $\mathrm{M}$ frames to receive a BW grant. If no BW is awarded during all of these frames then the BR is considered not successful and the SS again moves to the next contention round (departure from state  $(0, \mathrm{M})^{\mathrm{F}}$).

\begin{figure}[b]
	\centering
	\vspace{-18pt}
	\includegraphics[width=0.95\columnwidth]{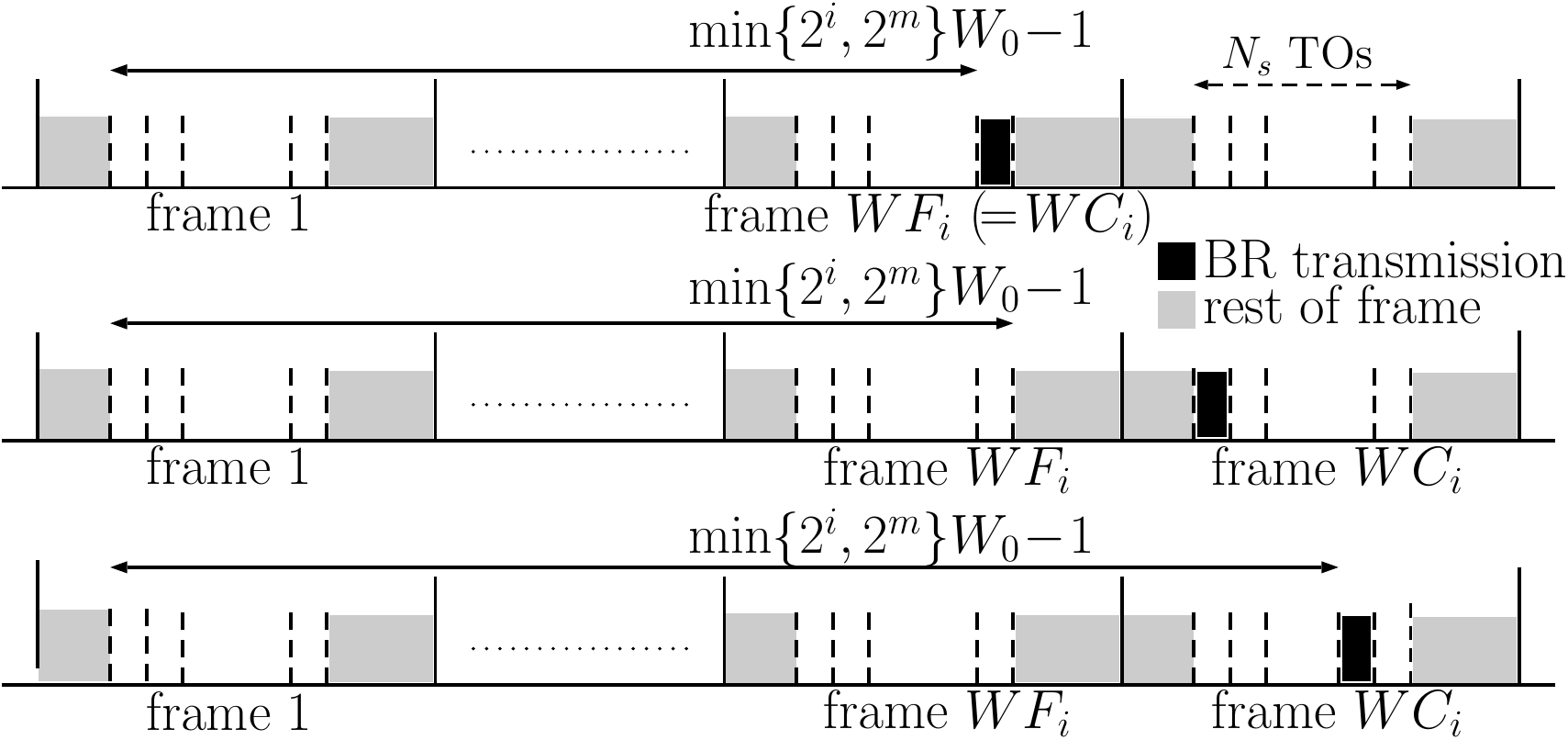}\\
	\caption{Maximum number of frames in a backoff period}\label{waiting-cases}
\end{figure}

The SS repeats the same process in the following contention round. In general, there are at maximum $\mathrm{D}+1$ rounds ($\mathrm{D}$ retransmissions) for completing a BR transmission and receiving a BW grant. After that, the packet corresponding to the BR is dropped. The only difference between successive rounds is the contention window used by the SS. Recall that when in contention round $i$, an SS will randomly choose the back-off counter $backoffcnt \in [0,2^{\min\{i,m\}}W_{0}-1]$ 
to indicate the number of TOs to wait before transmitting a BR. As a result, the BR will be transmitted in the $(backoffcnt\!+\!1)$-th consecutive TO. In general, this means that the SS
will wait on average $\overline{\mathrm{K}}_{i}$ frames before reaching state $(i,0)^{\mathrm{R}}$ because it is possible that $backoffcnt\!+\!1 > N_{s}$. 
To calculate $\overline{\mathrm{K}}_{i}$, let us model the number of waiting frames with a discrete random variable (RV) $\mathrm{K}_{i}$. Then, $\mathrm{J}_{i}=\mathrm{K}_{i}+1$ is also a RV that includes the frame in which the BR is actually transmitted, i.e. the frame corresponding to state $(i,0)^{\mathrm{R}}$. Observe that the sample space of $\mathrm{J}_{i}$ is $[1,\mathrm{WC}_{i}]$, where $\mathrm{WC}_{i}=\lceil2^{\min\{i,m\}}W_{0}/N_{s}\rceil$ (Fig.~\ref{waiting-cases}). Moreover, all values in $[1,\mathrm{WF}_{i}]$, where $\mathrm{WF}_{i}=\lfloor2^{\min\{i,m\}}W_{0}/N_{s}\rfloor$ and $\mathrm{WC}_{i}\!-\!1\!\leq\!\mathrm{WF}_{i}\!\leq\!\mathrm{WC}_{i}$, are equiprobable with probability $N_{s}/2^{\min\{i,m\}}W_{0}$. This is because for any value $x \in [1,\mathrm{WF}_{i}]$ there are exactly $N_{s}$ values of $backoffcnt$ that will result in  $\mathrm{J}_{i}=x$. In the case that $\mathrm{WC}_{i}=\mathrm{WF}_{i}\!+\!1$, i.e. when $2^{\min\{i,m\}}W_{0}/N_{s}$ is not an integer, $\mathrm{J}_{i}\!=\!\mathrm{WC}_{i}$ occurs with probability $(2^{\min\{i,m\}}W_{0}-\mathrm{WF}_{i}N_{s})/2^{\min\{i,m\}}W_{0}$. Consequently:
\begin{equation}\label{average_backoff}
\begin{split}
\overline{\mathrm{K}}_{i}\!& = \overline{\mathrm{J}}_{i}-1 =\!\!\sum_{j=1}^{\mathrm{WC}_{i}}\!j\mathrm{P}\{\mathrm{J}_{i}\!=\!j\} -1 \\ & =\frac{1}{2}\frac{\mathrm{WF}_{i}(\mathrm{WF}_{i}\!+\!1)N_{s}}{2^{\min\{i,m\}}W_{0}}
\!+\!\mathrm{WC}_{i}(1\!-\!\frac{\mathrm{WF}_{i}N_{s}}{2^{\min\{i,m\}}W_{0}})\!-\!1
\end{split}
\end{equation}

Let us now go back to the states $(i,j)^{\mathrm{T}}, \forall j \in [1,\mathrm{M}], i \in [0,\mathrm{D}]$, i.e. when the SS manages to successfully transmit data. The SS will now move to the idle state if no packet is waiting in the queue to be served. Let $\Pi_{0}$ denote this latter probability. Clearly, the SS will initiate the piggyback mechanism with probability $1\!-\!\Pi_{0}$ (transition to state $\mathrm{PG}_{1}$) and will continue to transmit data using the piggyback mechanism (states $\mathrm{PG}_{2}$ to $\mathrm{PG}_{G}$) if at the end of each transmission there is at least one packet waiting (probability $1\!-\!\Pi_{0}$). If after completing a piggyback transmission there is no waiting packet (probability $\Pi_{0}$) then the SS moves to the idle state. Note that the use of the piggyback mechanism is limited, therefore the SS, after $\mathrm{G}$ piggyback sessions, moves to the idle state with probability $\Pi_{0}$ (no packet is waiting) or enters contention to transmit a new BR (probability $1\!-\!\Pi_{0}$). To calculate $\Pi_{0}$, we model the SS queue as an M/G/1 queueing system with vacations\cite{kleinrock1975queueingsystems}, where packets are the customers and the service time is the total time that the SS is involved in transmitting a BR (including the backoff period) as well as the corresponding packet. Here, the vacation time is deterministic and equals $\mathrm{T_{fr}}$ because an SS with an empty queue at the start of a frame will pause serving packets arriving during this frame and will resume at the beginning of the next frame.
In the aforementioned system, the probability that a departing customer leaves no customers in the system is (see Appendix~\ref{appendix-mg1} for proof):
\begin{equation}\label{empty_queue}
\Pi_{0}=(1-\rho)\dfrac{1-e^{-\lambda\mathrm{T_{fr}}}}{\lambda\mathrm{T_{fr}}}
\end{equation}
where $\rho\!=\!\lambda\mathrm{E}\{S\}$, $\mathrm{E}\{S\}$ is the mean service time and $\rho_{in}\!=\!\lambda\mathrm{T_{fr}}$ the offered load. Note that (\ref{empty_queue}) holds for an arbitrary departing customer\cite{kleinrock1975queueingsystems},\cite{cooper1981queueingtheory}. Thus, $\Pi_{0}$ and the complementary probability $1\!-\!\Pi_{0}$ are the transition probabilities in any case of service completion and regardless of whether the packet is served using piggyback or the contention mechanism. For the same reason, we use $\Pi_{0}$ in the case of a dropped packet. Recall that this happens when the corresponding BR is not served after $D\!+\!1$ consecutive contention rounds (states $(\mathrm{D},\mathrm{M})^{\mathrm{C}}$ and $(\mathrm{D},\mathrm{M})^{\mathrm{F}}$). From the M/G/1 model's point of view, at that point the system concludes serving the packet therefore the SS moves to the idle state with probability $\Pi_{0}$ or enters a new contention round to serve a new waiting packet (probability $1\!-\!\Pi_{0}$). Finally, when the SS is in the idle state it will remain in that state with probability $p_{a}=e^{-\lambda\mathrm{T_{fr}}}$, i.e. if no packet arrival occurs during the duration of a frame, or will engage in contention otherwise.

\section{Performance Analysis of the Access Scheme}\label{network-analysis}

After modeling the procedure carried out by an SS we wish to use this model for analysing the performance of the access scheme at the network level. Note that the probability $b^{\mathrm{X}}_{i,j}$ of the SS being in state $(i,j)^{\mathrm{X}}$ of the proposed Markov chain can be expressed as a function of $b^{\mathrm{R}}_{0,0}$. The latter depends on three indicators (see Appendix~\ref{appendix-markov}) that portray the performance of the access scheme, namely: i) the collision probability $p$, ii) the mean service time $\mathrm{E}\{S\}$ and, iii) the probability $q$ of granting BW to a successfully transmitted contending BR. Our aim is to use the analysis of the access scheme at the network level so as to express each of the aforementioned indicators as a function of $b^{\mathrm{R}}_{0,0}$. In this way, it is possible to determine $p$, $\mathrm{E}\{S\}$ and $q$ by solving a system of three non-linear equations. 

\subsection{Collision Probability}\label{p-derivation}

Let $P_{\mathrm{B}}$ denote the probability that an SS will try to transmit a BR in a specific frame. Clearly, this is the probability that the SS will transmit a BR in any contention round, i.e. the sum of the probabilities of the SS being in states $(i,0)^{\mathrm{R}}, i\in [0,\mathrm{D}]$. Therefore (for proof refer to (\ref{trans_slot}), Appendix~\ref{appendix-markov}):
\begin{equation}\label{prob-bwreq}
P_{\mathrm{B}} = \sum_{i=0}^{\mathrm{D}}b_{i,0}^{\mathrm{R}} = \dfrac{1-(p_{f})^{\mathrm{D}+1}}{1-p_{f}} b_{0,0}^{\mathrm{R}}=\tau b_{0,0}^{\mathrm{R}}
\end{equation}
where $p_{f}=p+(1-p)(1-q)^{\mathrm{M}}$. Provided that an SS transmits a BR, the probability of not being involved in a collision ($1\!-\!p$) is the probability that none of the remaining $N\!-\!1$ subscribers will try to transmit in the same TO. Clearly, the number of subscribers trying to transmit in a specific TO can be modeled as a binomial RV. Therefore,
\begin{equation}\label{p_col}
\begin{split} 
p &= 1-\mathrm{P}\{\text{no transmiting SS out of } N-1\} \\
&= 1\!-\!{(1-\frac{P_{\mathrm{B}}}{N_s})}^{N-1}\!=\!1\!-\!{(1-\frac{\tau}{N_s} b_{0,0}^{\mathrm{R}})}^{N-1}
\end{split} 
\end{equation}
where we have also assumed that it is equiprobable for an SS to select any TO.  

\subsection{Mean Service Delay}\label{S-derivation}
	
There are two alternatives for serving a packet, i.e. either through a contending or a piggybacked BR. 
In the case of contention, a BR will be successful and a packet will be transmitted if the SS ends up in one of the states $(i,j)^{\mathrm{T}}, \forall i\in [0,\mathrm{D}], j \in [1,\mathrm{M}]$. The probability of reaching state $(i,j)^{\mathrm{T}}$ provided that an SS creates a BR can be written (using (\ref{success_and_fail})):
\begin{equation}
\begin{split}
bb_{i,j} &= \mathrm{P}\{\textrm{SS in } (i,j)^{\mathrm{T}} |\textrm{ SS creates a BR}\}\\
&=\dfrac{b_{i,j}^{\mathrm{T}}}{b_{0,0}^{\mathrm{R}}}=q(1\!-\!p)(1\!-\!q)^{j-1}p_{f}^{i}
\end{split}
\end{equation}
Using the Markov chain we can find that the time elapsed from the start of contention until the SS reaches state $(i,j)^{T}$ is (expressed in frames):
\vspace{-3pt}
\begin{align}\label{delay_to_T}
  \mathrm{E}\{S_{i,j}\} & =j+i\mathrm{M}+\sum_{k=0}^{i}\overline{\mathrm{K}}_{k}, & \forall j\in [1,\mathrm{M}], i\in [0,\mathrm{D}]
\end{align}
The packet is dropped if the SS ends up in one of the states $(\mathrm{D},\mathrm{M})^{\mathrm{C}}$ and $(\mathrm{D},\mathrm{M})^{\mathrm{F}}$. This happens with probability
\begin{equation}\label{pdrop}
\begin{split}
	P_{\mathrm{D}}&=\mathrm{P}\{\textrm{packet dropped }|\textrm{ SS creates a BR}\}\\ &=\dfrac{b_{\mathrm{D},\mathrm{M}}^{\mathrm{C}}+b_{\mathrm{D},\mathrm{M}}^{\mathrm{F}}}{b_{0,0}^{\mathrm{R}}}=(p_{f})^{\mathrm{D}+1}
\end{split}
\end{equation}
while the time spent is:
\vspace{-3pt}
\begin{align}\label{delay_to_D}
  \mathrm{E}\{S_{\mathrm{D}}\} & =(\mathrm{D}+1)\mathrm{M}+\sum_{k=0}^{\mathrm{D}}\overline{\mathrm{K}}_{k}.
\vspace{-3pt}  
\end{align}
As a result, the expected service time for a packet served through the contention mechanism is:
\begin{equation}\label{delay_contention}
\mathrm{E}\{S_{\mathrm{C}}\}=(p_{f})^{\mathrm{D}+1}\mathrm{E}\{S_{\mathrm{D}}\} + \sum_{i=0}^{\mathrm{D}}\sum_{j=1}^{\mathrm{M}} bb_{i,j} \mathrm{E}\{S_{i,j}\} 
\end{equation}

For packets served through the piggyback mechanism the service delay is clearly equal to one frame, i.e. $S_{\mathrm{P}}\!=\!1$. Note that each packet transmitted using the contention mechanism is followed on average by $\bar{G}$ packet transmissions enabled through piggybacked BRs. Therefore the overall expected service delay is 
\begin{equation}\label{mean_service_delay-first}
\mathrm{E}\{S\} = \dfrac{\mathrm{E}\{S_{\mathrm{C}}\}+\bar{G}\mathrm{E}\{S_{\mathrm{P}}\}}{1+\bar{G}} \nonumber
\end{equation}
Since in each piggyback state $\mathrm{PG}_{i}$ corresponds one data packet then:
\begin{equation}\label{avg-g}
\begin{split}
\bar{G}\!=\!\!\sum_{i=1}^{\mathrm{G}}\!\mathrm{\mathrm{P}}\{\mathrm{PG}_{i} | \textrm{SS creates a BR}\}
\!=\!\!\sum_{i=1}^{\mathrm{G}}\!\dfrac{\mathrm{\mathrm{P}}\{\mathrm{PG}_{i}\}}{b_{0,0}^{\mathrm{R}}}\!=\!Z
\end{split}
\end{equation}
where we used (\ref{piggy_states}). As a result,
\begin{equation}\label{mean_service_delay}
\begin{split}
\mathrm{E}\{S\}\!&=\!\dfrac{\mathrm{E}\{S_{\mathrm{C}}\}\!+\!Z}{1+Z}\\
&=\dfrac{(p_{f})^{\mathrm{D}\!+\!1}\mathrm{E}\{S_{\mathrm{D}}\}\!\!+\!\!\sum_{i=0}^{\mathrm{D}}\!\sum_{j=1}^{\mathrm{M}} \!bb_{i,j} \mathrm{E}\{S_{i,j}\}\!+\!Z}{1+Z}
\end{split}
\end{equation}

\subsection{Probability of Bandwidth Allocation}\label{q-derivation}

Assume that we examine a successful BR that awaits BW allocation in a specific frame and let $\mathcal{R}$ denote the total number of BRs awaiting BW allocation. Also let $\mathcal{Q}$ denote the same number excluding the examined BR, i.e. $\mathcal{R}=\mathcal{Q}+1$. Also, let $\mathcal{P}$ denote the number of data slots to be allocated to piggybacked BRs. Recall that piggybacked BRs are served with priority over contending BRs. This policy is necessary for guaranteeing bandwidth allocation to piggybacked BRs. Therefore, $\mathcal{P}$ data slots are allocated to the SSs that are in the piggyback mode while $L\!-\!\mathcal{P}$ remaining data slots are to be allocated to $\mathcal{R}$ pending BRs. Clearly, the event of not serving some of the pending BRs occurs when $\mathcal{R}\!>\!L-\mathcal{P}$ with $\mathcal{R}\!-\!(L-\mathcal{P})$ out of the $\mathcal{R}$ BRs not receiving BW. Let $I_{b}$ be the indicator RV associated with the event of not allocating BW to the examined BR. Under the assumption that all BRs are served with the same probability,
\vspace{-3pt}
\begin{equation}\label{q_helper}
\mathrm{P}\{I_{b}=1|\mathcal{R}=k,\mathcal{P}=l\} = \frac{k\!-\!(L\!-\!l)}{k}
\vspace{-1pt}
\end{equation}
i.e., this is the probability that the examined BR is one of the $k-(L-l)$ that will not be served out of $k$ pending BRs. 
Then, the probability $r\!=\!1\!-\!q$ of not granting BW to the examined BR is
\begin{equation}\label{q_basic}
\begin{split}
r &= \mathrm{P}\{I_{b}=1|\mathcal{R},\mathcal{P}\}\mathrm{P}\{\mathcal{R}\!>\!L\!-\!\mathcal{P}\}\\
&=\!\sum_{i}\mathrm{P}\{I_{b}\!=\!1|\mathcal{R},\mathcal{P}\!=\!i\}\mathrm{P}\{\mathcal{R}\!>\!L\!-\!i\}\mathrm{P}\{\mathcal{P}\!=\!i\}\\
&=\!\sum_{i}\mathrm{P}\{I_{b}\!=\!1|\mathcal{R},\mathcal{P}\!=\!i\}\mathrm{P}\{\mathcal{Q}\!+\!1\!>\!L\!-\!i\}\mathrm{P}\{\mathcal{P}\!=\!i\}\\
&=\sum_{i}\sum_{j>L-i-1}\Big[\mathrm{P}\{I_{b}\!=\!1|\mathcal{R}\!=\!j+1,\!\mathcal{P}\!=\!i\}\\
&\qquad \qquad \;\;\;\;\;\; \mathrm{P}\{\mathcal{Q}\!=\!j|\mathcal{P}\!=\!i\}\mathrm{P}\{\mathcal{P}\!=\!i\}\Big]
\end{split}
\vspace{-6pt}
\end{equation}

In order to determine $r$ (or equivalently $q$) we should determine both $\mathrm{P}\{\mathcal{P}\!=\!i\}$ and $\mathrm{P}\{\mathcal{Q}\!=\!j|\mathcal{P}\!=\!i\}$. It is possible to model the allocation of data slots to the piggyback mechanism using a set of $L$ independent Bernoulli RVs. In this case, we can model $\mathcal{P}$ as a binomial RV, therefore
\vspace{-3pt}
\begin{equation}\label{piggyback_dist}
\mathrm{P}\{\mathcal{P}=i\}=\binom{L}{i}\mathrm{P}^{i}_{G}(1-P_{\mathrm{G}})^{L-i}
\vspace{-3pt}
\end{equation}
where $P_{\mathrm{G}}=\mathrm{E}\{\mathcal{P}\}/L$ is the probability that a data slot is allocated to piggybacking. Observe that in any given frame an SS has either one pending or a piggybacked BR. Therefore, if $\mathcal{P}=i$ SSs are in piggyback mode and one SS has already produced the examined BR then there are $N-i-1$ SSs that may have a pending BR. As a result, we can model $\mathcal{Q}$ as a sum of $N-i-1$ Bernoulli RVs. Therefore the conditional pmf of $\mathcal{Q}$ given $\mathcal{P}$ is Binomial, i.e.
\begin{equation}\label{req_dist}
\mathrm{P}\{\mathcal{Q}=j|\mathcal{P}=i\}=\binom{N\!-\!i\!-\!1}{j}\mathrm{P}^{j}_{A}(1-P_{\mathrm{A}})^{N-i-1-j}
\end{equation}
where $P_{\mathrm{A}}=\mathrm{E}\{\mathcal{R}\}/(N-i)$ is the probability of an SS that is not in piggyback mode to have a pending BR. By combining (\ref{q_basic}) with (\ref{q_helper}),(\ref{piggyback_dist}) and (\ref{req_dist}) we conclude that
\begin{equation}\label{q_derivation}
\begin{split}
r &\!=\!\!\sum_{i=0}^{L}\!\sum_{\substack{j=L-i}}^{N-i-1}\!\!\Bigg[\!\frac{j\!+\!1\!-\!(L\!-\!i)}{j\!+\!1}\!\binom{N\!-\!i\!-\!1}{j}P_{\mathrm{A}}^{j}\\
&\qquad\qquad\qquad (1\!\!-\!\!P_{\mathrm{A}})^{N-i-1-j}\binom{L}{i}P_{\mathrm{G}}^{i}(1\!\!-\!\!P_{\mathrm{G}})^{L\!-i} \Bigg]
\end{split}
\end{equation}
Note that it is possible to derive $q$ using (\ref{q_derivation}) as long as we determine both $P_{\mathrm{G}}$ and $P_{\mathrm{A}}$ or equivalently $\mathrm{E}\{\mathcal{P}\}$ and $\mathrm{E}\{\mathcal{R}\}$. In Appendix~\ref{appendix-q-related} we prove that:
\begin{equation}\label{er-formula}
\begin{split}
\mathrm{E}\{\mathcal{R}\}=N_{s}P_{\mathrm{S}}\sum_{i=0}^{\mathrm{M}\!-\!1}(1-q)^{i}
\end{split}	
\end{equation}
and
\vspace{-6pt}
\begin{equation}\label{ep-formula}
\begin{split}
\mathrm{E}\{\mathcal{P}\}&=\mathrm{E}\{\mathcal{R}\}q\sum_{i=0}^{\mathrm{G}-1}(1-\Pi_{0})^{i+1}
\end{split}	
\end{equation}
where $P_{\mathrm{S}}$ is the probability of a successful BR transmission in a TO. In order to calculate $P_{\mathrm{S}}$ observe that a successful BR is one not involved in a collision. In other words, a TO contains a successful BR if only one SS transmits in the slot, therefore,  
\begin{equation}\label{ps_value}
P_{\mathrm{S}}\!=\!N\dfrac{P_{\mathrm{B}}}{N_{s}}(1\!-\!\dfrac{P_{\mathrm{B}}}{N_{s}})^{N-1}\!=\!N\dfrac{\tau b^{\mathrm{T}}_{0,0}}{N_{s}} (1\!-\!\dfrac{\tau b^{\mathrm{T}}_{0,0}}{N_{s}})^{N-1}
\end{equation}
where $P_{\mathrm{B}}$ is the probability that an SS transmits a BR in a frame and is given by (\ref{prob-bwreq}).

\subsection{Determining System Performance}\label{more_metrics}

Up to this point we have managed to come up with three equations that involve only three variables. More specifically, observe that equations (\ref{p_col}), (\ref{mean_service_delay}) and (\ref{q_derivation}) form a system of three equations where only the performance indicators $p$, $\mathrm{E}\{S\}$ and $q$ are unknown. Therefore, it is possible to numerically solve this system and determine these indicators as well as other performance metrics that up to now we have managed to express as a function of these indicators such $P_{\mathrm{S}}$, $\bar{G}$, $P_{\mathrm{D}}$, etc. It is also possible to extend our analysis in order to explore more performance aspects of the access scheme. One such interesting aspect is the overall achieved throughput. Recall that the probability of a TO containing a successful BR is $P_{\mathrm{S}}$ and can be calculated using (\ref{ps_value}). At the same time, a BW grant may be allocated to a successful BR in one of the  $\mathrm{M}$ consecutive frames of the T16 period with probability $q$ per frame. Therefore, the probability that a successful BR is actually served in a contention round is $q_{_{\mathrm{M}}}=1\!-\!(1\!-\!q)^{\mathrm{M}}$. Consequently, the throughput achieved through contention, expressed in packets per frame, is:  
\begin{equation}\label{throughput}
Th_{\mathrm{C}}=[1-(1-q)^{\mathrm{M}}]P_{\mathrm{S}}N_{s}= q_{_{\mathrm{M}}}P_{\mathrm{S}}N_{s}
\end{equation}
To calculate the throughput achieved by means of the piggyback mechanism bear in mind that each attempt from an SS to transmit a contending BR (which occurs with probability $b^{\mathrm{R}}_{0,0}$) finally succeeds with probability $1\!-\!P_{\mathrm{D}}$. Furthermore, each packet sent after a successful BR is followed on average by $\bar{G}\!=\!Z$ packets sent using piggybacked BRs. As a result, the throughput of the piggyback mechanism is:
\begin{equation}\label{throughput-pg}
Th_{\mathrm{PG}}=Z N (1-P_{\mathrm{D}}) b^{\mathrm{R}}_{0,0}
\end{equation}
where $Z$, $P_{\mathrm{D}}$ and $ b^{\mathrm{R}}_{0,0}$ are given by (\ref{avg-g}), (\ref{pdrop}) and (\ref{normalization}) respectively. The overall achieved throughput is obviously $Th=Th_{\mathrm{C}}+Th_{\mathrm{PG}}$.

Another very interesting performance metric is the time that an arriving packet has to wait before being served. As we mentioned earlier, we model an SS's queue as an M/G/1 system with vacations. Therefore it is reasonable to consider the Pollaczek-Khinchine formula as a starting point. However, in our system the service delay depends on whether an arriving packet enters the queue or not. In Appendix~\ref{appendix-waiting-delay} we approximate the expected waiting delay as:
\begin{equation}\label{waiting-delay}
	\mathrm{E}\{W\}\simeq\frac{\lambda\mathrm{E}\{S^{2}\}}{2(1-\lambda\mathrm{E}\{S_{Q}\})}+\frac{\mathrm{T_{fr}}}{2}
\end{equation}
where $\mathrm{E}\{S_{Q}\}$ is the expected service time seen by a queued packet and is given by (\ref{eq:inqueue-delay}), while $\mathrm{E}\{S^{2}\}$ is the second order moment of the expected service delay and can be calculated using (\ref{mean_square_service_delay}). Note that (\ref{waiting-delay}) can be seen as a modified version of the Pollaczek-Khinchine formula that results if we replace $\mathrm{E}\{S\}$ with $\mathrm{E}\{S_{Q}\}$ in the original formula.

\section{Results \& Model Validation}\label{evaluation}

\begin{table}[!b]
	\vspace{-18pt}
	\centering
	\caption{Parameters used in Simulations}
	\vspace{-6pt}
	\label{sim_params}	
	\begin{tabu}{|[0.8pt]c|c|c|c|c|[0.8pt]} 
		\tabucline[0.8pt]{-}
		\multicolumn{5}{|[0.8pt]c|[0.8pt]}{Range of parameters}\\
		\hline
		\multirow{ 2}{*}{$N \in [2,50]$} & \multirow{ 2}{*}{$L \in [5,21]$} & \multicolumn{2}{c|}{$\lambda \in [0.05, 2]$ packets/frame} & \multirow{ 2}{*}{$\mathrm{G} \in [0,5]$} \\
		& & \multicolumn{2}{c|}{$\rho_{in}\in [0.05, 1]$ Erlang} & \\
		\hline
		\multicolumn{5}{|[0.8pt]c|[0.8pt]}{Default values}\\
		\hline
		$N_{s}=20$ & $W_{0}=32$ & $m=5$ & $\mathrm{M}=6$ & $\mathrm{D}=5$\\
		\tabucline[0.8pt]{-}
	\end{tabu}
\end{table}
\begin{figure*}[!t]
	{\centering
		\captionsetup[subfloat]{farskip=2pt,captionskip=1pt}
		\subfloat[Case I][]
		{
			\includegraphics[width=0.33\textwidth]{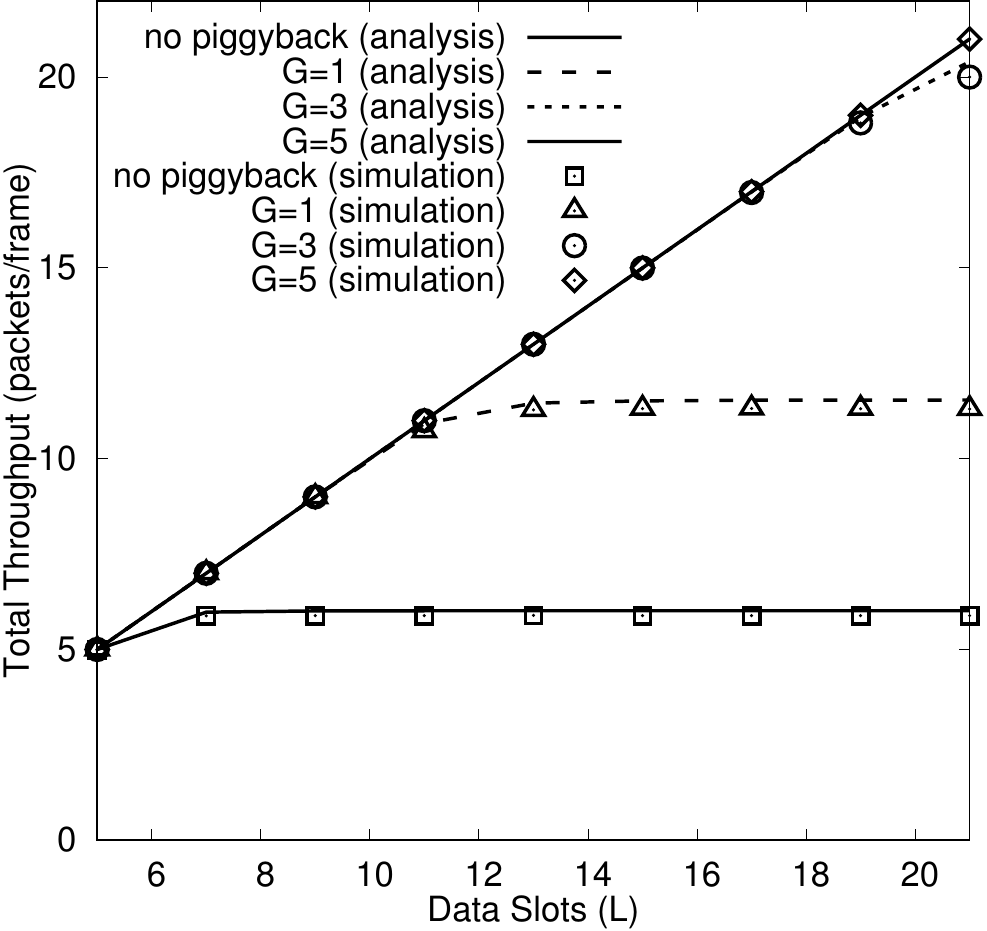}
			\label{Thput_vs_CH}
		}
		\subfloat[Case II][]
		{
			\includegraphics[width=0.33\textwidth]{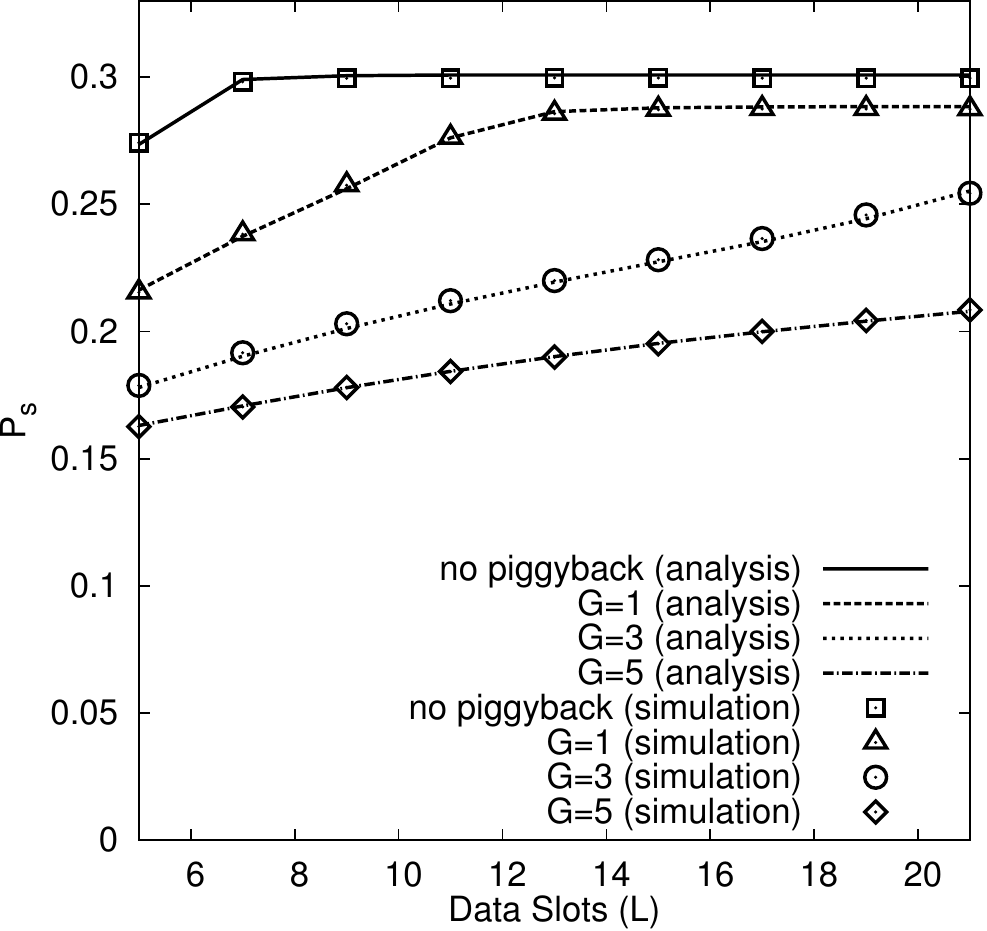}
			\label{Ps_vs_CH}
		}
		\subfloat[Case III][]
		{
			\includegraphics[width=0.33\textwidth]{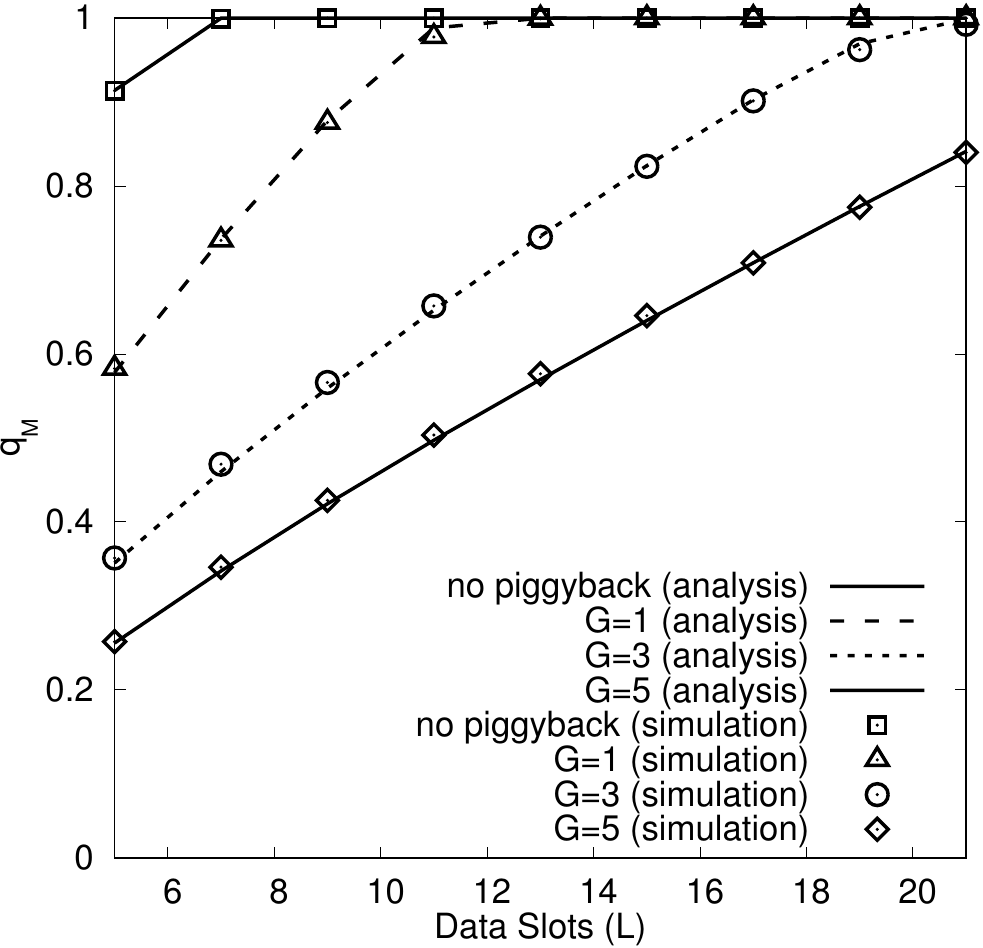}
			\label{q_vs_CH}
		}
		\\ \hfil%
		\subfloat[Case IV][]
		{
			\includegraphics[width=0.42\textwidth]{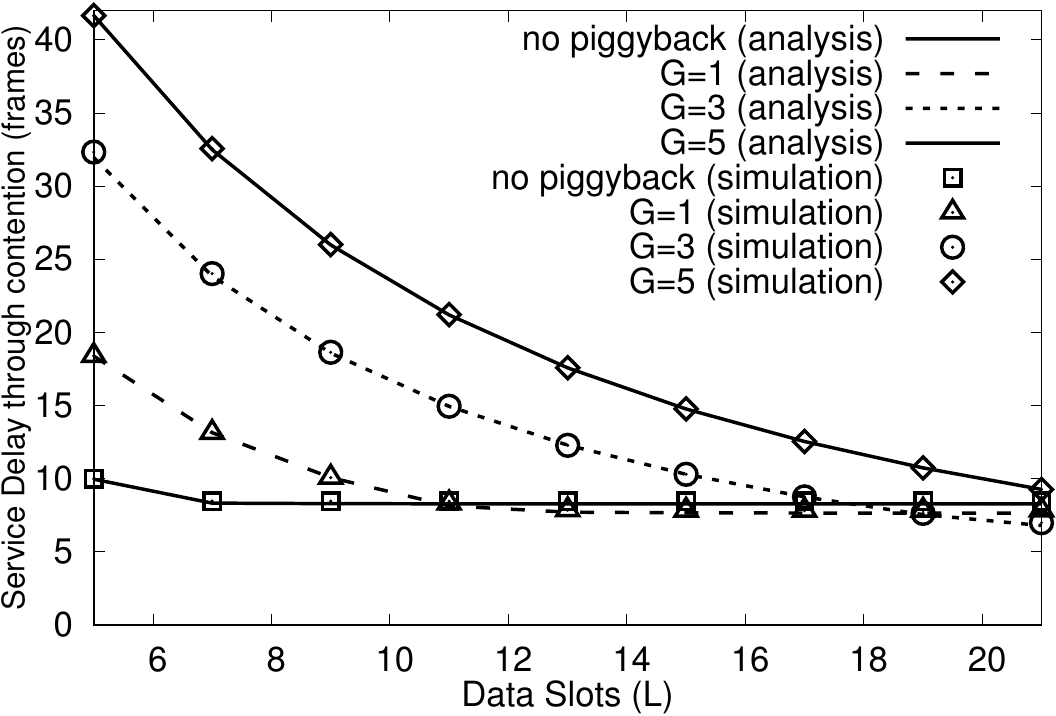}
			\label{Acc_vs_CH}
		}
		\hfil
		\subfloat[Case V][]
		{
			\includegraphics[width=0.42\textwidth]{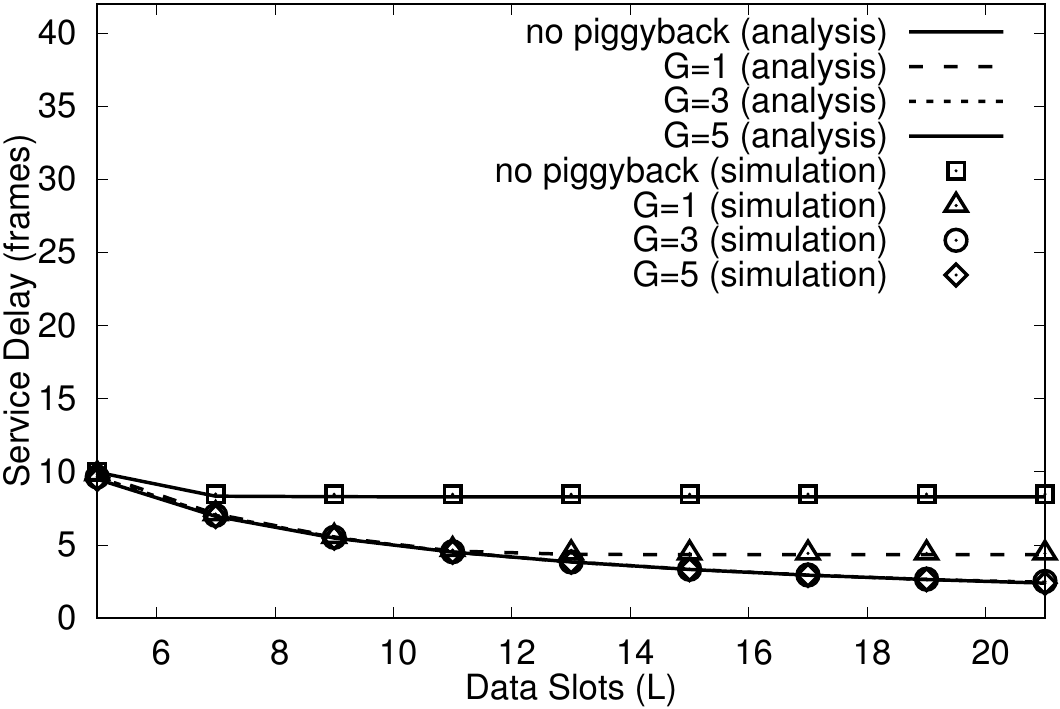}
			\label{ET_vs_CH}
		}			
	}
	\vspace{-3pt}
	\caption{Performance vs number of data slots ($L$) under saturated conditions ($\rho_{in}=1$): (a) Total Throughput ($\mathrm{Th}$) (b) $P_{\mathrm{S}}$, (c) $q$, (d) Service delay through contention ($\mathrm{E}\{S_{\mathrm{C}}\}$) and (e) Service delay ($\mathrm{E}\{S\}$)}
	\label{Per_vs_CH_1}
	\vspace{-12pt}
\end{figure*}
In order to investigate the impact of the piggyback mechanism on the IEEE 802.16 contention-based access scheme as well as to verify the proposed analytical model we conducted several simulation experiments. To this end, we used a custom event-driven simulator, written in C++, that implements broadcast polling with the piggyback mechanism as specified in the IEEE 802.16 standard\cite{ieee2009ieee}. 
Table~\ref{sim_params} presents the parameters used in our simulations. Since we focus on exploring the features of piggybacking, we refrain from investigating the impact of parameters associated with broadcast polling such as $W_{0}$, $N_{s}$, etc. The role of such parameters has been extensively investigated in the literature~\cite{fallah-model},\cite{giambene2013nonsaturated}. Therefore we choose typical default values based on this literature. In any case, the interested reader can refer to~\cite{TR-2016-2} where we present a more extensive set of experiments, including ones exploring the impact of the aforementioned parameters. Note that we use $N_{s}\!=\!20\!<\!W_{0}$. In this case the probability that an SS transmits a BR is not the same for all TOs. However, we make this choice to illustrate the accuracy of our model even in such a challenging scenario.
Finally, to provide a more generic analysis we use the frame duration as the time unit. For example, throughput related metrics are expressed on a per frame basis. Similarly, the unit for delay related metrics is the frame duration.  

In the first experiment we focus on capturing the full extent of the benefits associated with the piggyback mode. In particular, we investigate the performance of the piggyback-enabled random access method versus the number of data slots $L$ in saturation conditions, i.e. $\rho_{in}\!=\!\lambda\mathrm{T_{fr}}\!=\!1$. Fig.~\ref{Thput_vs_CH} illustrates the overall throughput (attained either through contention or piggyback) for different values of $\mathrm{G}$. Observe that when piggyback is disabled (i.e. $\mathrm{G}\!=\!0$) the saturation throughput is achieved for $L\!=\!7\!=\!\lceil0.33N_{s}\rceil$. This result has previously been found in \cite{vinel2009capacity} where the authors prove that the optimal data slots to TOs ratio is $\frac{L}{N_{s}}\!=\!\frac{\ln2}{2}\!\approx\!1/3$. Note that without piggyback the saturation throughput is achieved when the probability of successful BR transmission ($P_{\mathrm{S}}$) is maximized (Fig.~\ref{Ps_vs_CH}). When $L\!<\!7$ the throughput decreases since the available data slots are not sufficient for satisfying the offered load. This is confirmed by Fig.~\ref{q_vs_CH} where $q_{_{\mathrm{M}}}\!<\!1$ when $L\!<\!7$, i.e. it is possible for a successful BR not to receive a BW grant even if it waits for $\mathrm{M}$ frames. 

On the other hand, when piggyback is enabled ($\mathrm{G}\!>\!0$) the number of data slots required for maximizing throughput is clearly greater than $0.33N_{s}$. This is reasonable if we bear in mind that each BW grant allocated to a successful BR is followed on average by the transmission of $\bar{G}$ packets through piggybacking. Thus, apart from the BW allocated to successful BRs, BW is also required for the piggyback mechanism. A rough estimation for the number of data slots required for maximizing throughput in saturation conditions is given by
\begin{equation}
	L'(\mathrm{G})=\lceil(\mathrm{G}+1)\mathrm{P_{S}^{max}}N_{s}\rceil\label{eq:maxL}
	\vspace{-4pt}
\end{equation}
where $\mathrm{P_{S}^{max}}$ is the maximum probability of successful BR transmission. We also use $\mathrm{G}$ instead of $\bar{G}$ because in saturation $\bar{G}\!=\!\mathrm{G}$. Observe that $\mathrm{P_{S}^{max}}N_{s}$ is actually the average number of successful BRs while $\mathrm{GP_{S}^{max}}N_{s}$ is the BW required for packets sent using piggybacked BRs. Similarly, the maximum throughput (in pkts/frame) can be approximated as 
\begin{equation}
	Th^{\mathrm{MAX}}=\min\{L,L'(\mathrm{G})\}\!=\!\min\{L,(\mathrm{G}\!+\!1)\mathrm{P_{S}^{max}}N_{s}\}\label{eq:maxthapprox}
\end{equation}
In general, when $L\!<\!L'(\mathrm{G})$ throughput is upper bounded by $L$ and no further improvement is possible regardless of whether we increase $\mathrm{G}$ because it is not possible to exceed $L$ pkts/frame. For example, as illustrated in Fig.~\ref{Thput_vs_CH}, when $\mathrm{G}\!=\!1$ the throughput is upper bounded by $L$ for every value $L\!<\!L'(1)\!=\!12$. Therefore, there is no point in using $\mathrm{G\!>\!1}$ when $L\!<\!12$ because in all cases the throughput would not exceed $L$ pkts/frame (note that performance is identical for all $\mathrm{G}\!>\!0$ when $L\!<\!12$). To explain this result from another point of view recall that in each frame $\mathrm{E}\{\mathcal{P}\}$ data slots are occupied by the piggyback mechanism and only $L\!-\!\mathrm{E}\{\mathcal{P}\}$ are available for successful contending BRs. As $\mathrm{G}$ increases $L\!-\!\mathrm{E}\{\mathcal{P}\}$ is cut back. Consequently, not only it is less probable that a BR will receive a grant (Fig.~\ref{q_vs_CH}) but also fewer SSs are involved in transmitting new contending BRs (observe the reduction of $P_{\mathrm{S}}$ in Fig.~\ref{Ps_vs_CH}). The latter is because more SSs are entangled in long waiting periods in order to receive a BW grant and therefore they become idle. Overall, fewer packets are transmitted through contention. This reduces the piggyback throughput since piggybacking is only possible after the transmission of a packet through contention. In order to break this negative feedback cycle and improve throughput we need to increase both $L$ and $\mathrm{G}$. In this way, we avoid reducing $L\!-\!\mathrm{E}\{\mathcal{P}\}$ excessively and therefore allow $q_{_{\mathrm{M}}}$ to increase. In other words, to obtain a throughput gain we should increase $L$ beyond $L'(1)\!=\!12$ and then use $\mathrm{G}\!>\!1$.

Going back to the case that $L\!\geq\!L'(\mathrm{G})$, it is easy to show (by combining (\ref{mean-p-calligr}) and (\ref{mean-r-calligr})) that in saturation $L\!-\!\mathrm{E}\{\mathcal{P}\}\!>\!P_{\mathrm{S}}N_{s}$, i.e. there is always enough bandwidth to serve all successful BRs (i.e. $q_{_{\mathrm{M}}}\!=\!1$), therefore it is possible to maximize throughput. This is the case for example when $\mathrm{G}\!=\!1$ and $L\!\geq\!L'(1)\!=\!12$ as can be seen in Fig.~\ref{Thput_vs_CH}-\ref{q_vs_CH}. However, note that the upper bound of throughput does not increase linearly with $\mathrm{G}$. For example, for $G\!=\!1$ the upper bound is $\sim\!11.3$ pkts/frame while for $G\!=\!3$ it is only $\sim\!20$ pkts/frame. According to (\ref{eq:maxthapprox}) the upper bound depends on $\mathrm{G}$, $\mathrm{P_{S}^{max}}$ and $N_{s}$. Interestingly, for the reasons explained previously, $\mathrm{P_{S}^{max}}$ itself depends on $\mathrm{G}$, i.e. for a certain $L$ value $\mathrm{P_{S}^{max}}$ decreases as $\mathrm{G}$ increases (Fig.~\ref{Ps_vs_CH}). This explains the non-linear dependency of maximum throughput and $\mathrm{G}$.

In Fig.~\ref{Acc_vs_CH} we present the service delay for packets transmitted through the contention mechanism ($\mathrm{E}\{S_{C}\}$) while Fig.~\ref{ET_vs_CH} illustrates the service delay for all transmitted packets ($\mathrm{E}\{S\}$). Evidently, $\mathrm{E}\{S_{C}\}$ is higher when $\mathrm{G}\!>\!0$ compared to the case that piggyback is disabled unless a considerable amount of bandwidth is available. This is a direct indication of the more severe competition for available bandwidth when the latter is not in abundance. Indeed, as mentioned previously, when $\mathrm{G}\!>\!0$ only $L\!-\!\mathrm{E}\{\mathcal{P}\}$ data slots are available for successfully contending BRs. Hence, it is more difficult to receive a BW grant in a specific frame (Fig.~\ref{q_vs_CH}). As a result, on average more frames are required for receiving a BW grant, thus increasing service delay. The situation is reversed when bandwidth is not an issue (see Fig.~\ref{Acc_vs_CH}, $L\!=\!21$). In such cases the piggyback mechanism takes full advantage of bandwidth availability and significantly reduces the delay. We provide more insight into this performance aspect in the next experiment. Another important observation is that $\mathrm{E}\{S_{C}\}$ increases dramatically for smaller values of $L$ as well as for greater values of $\mathrm{G}$ since in both cases the $L\!-\!\mathrm{E}\{\mathcal{P}\}$ available data slots are reduced dramatically. Nevertheless, increasing $\mathrm{G}$ has a positive impact on the overall service delay $\mathrm{E}\{S\}$. The reason is that more packets are transmitted with minimum delay using the piggyback mechanism. In accordance with our observations so far, obtaining delay gains requires increasing both $\mathrm{G}$ and $L$ at the same time. Nonetheless, using piggyback to reduce delay involves also a downside. The delay jitter for two successive packets increases when one packet is sent using piggybacking while the other uses a contending BR.

\begin{figure*}[!tbp]
	{\centering
		\captionsetup[subfloat]{farskip=2pt,captionskip=1pt}
		\subfloat[Case I][]
		{
			\includegraphics[width=0.325\textwidth]{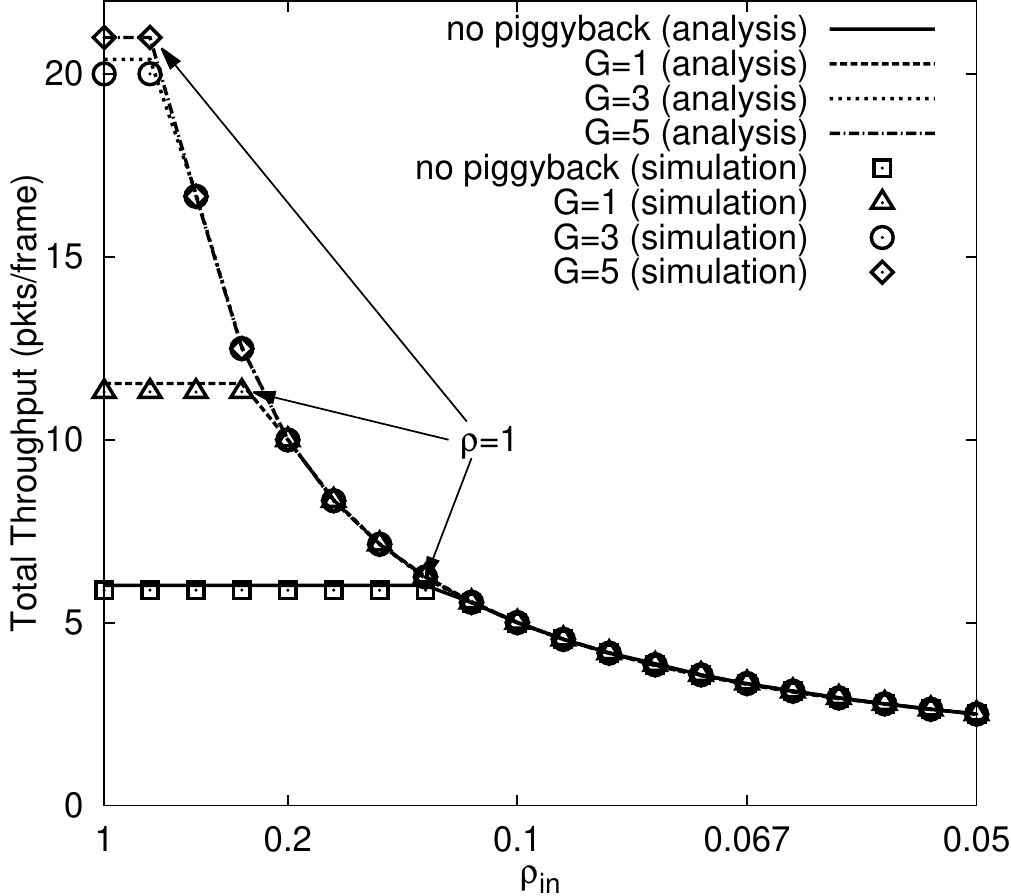}
			\label{Thput_vs_load_CH21}
		}
		\subfloat[Case II][]
		{
			\includegraphics[width=0.325\textwidth]{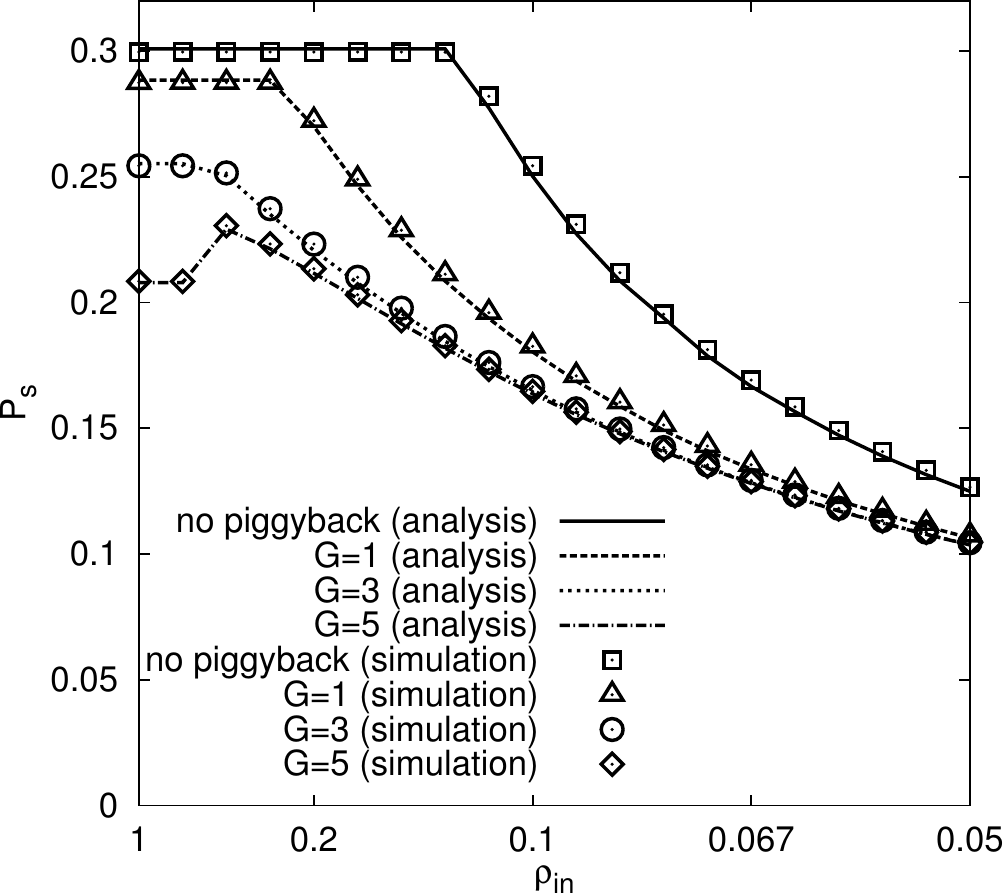}
			\label{Ps_vs_load_CH21}
		}
		\subfloat[Case III][]
		{
			\includegraphics[width=0.325\textwidth]{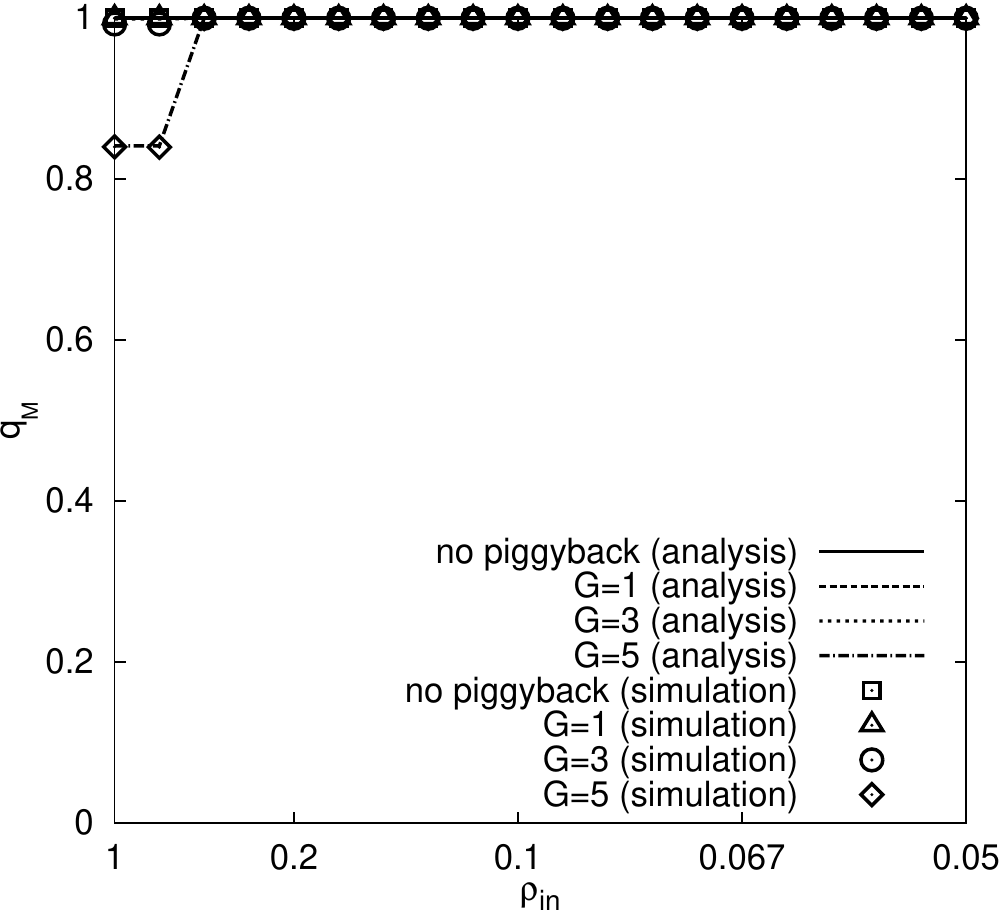}
			\label{q_vs_load_CH21}
		}	
		\\
		\subfloat[Case IV][]
		{
			\includegraphics[width=0.325\textwidth]{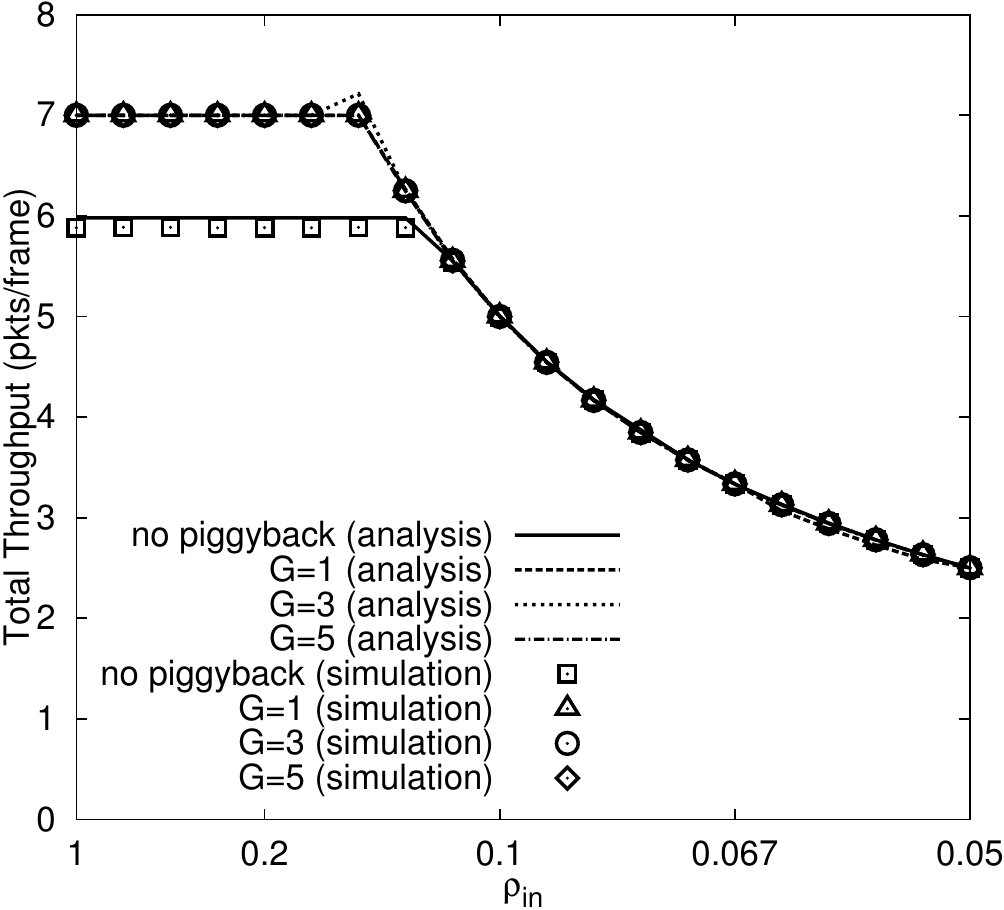}
			\label{Thput_vs_load_CH7}
		}
		\subfloat[Case V][]
		{
			\includegraphics[width=0.325\textwidth]{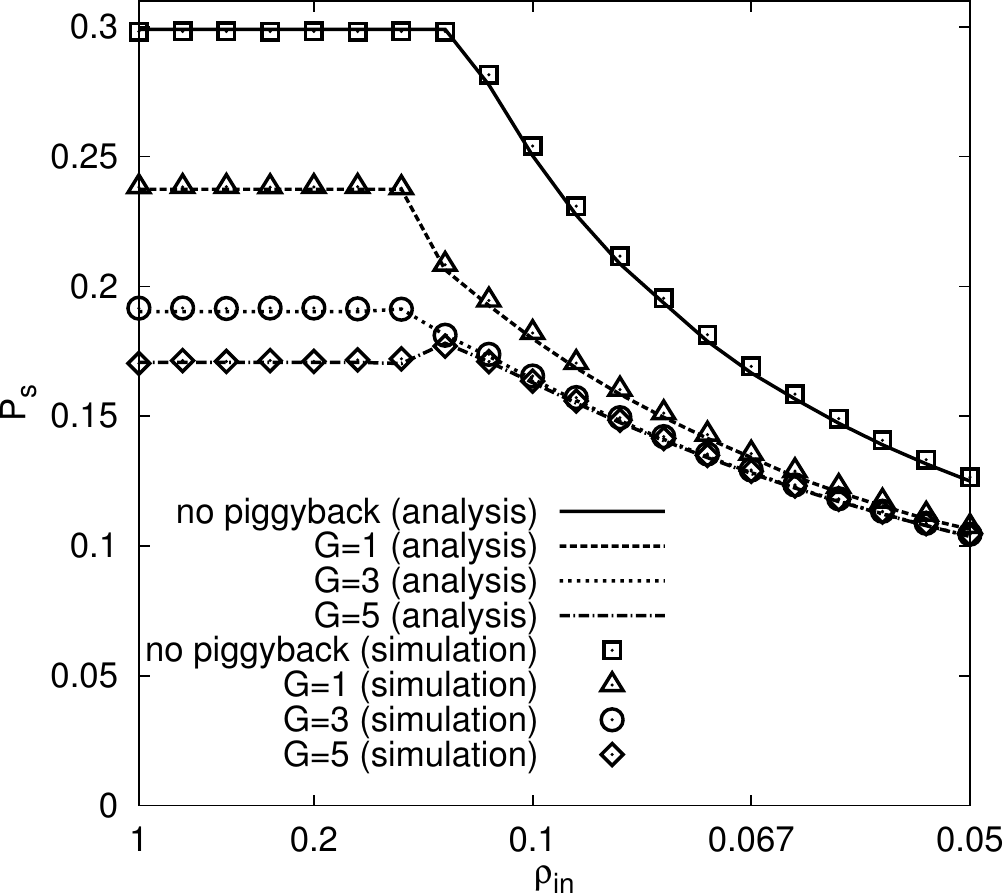}
			\label{Ps_vs_load_CH7}
		}
		\subfloat[Case VI][]
		{
			\includegraphics[width=0.325\textwidth]{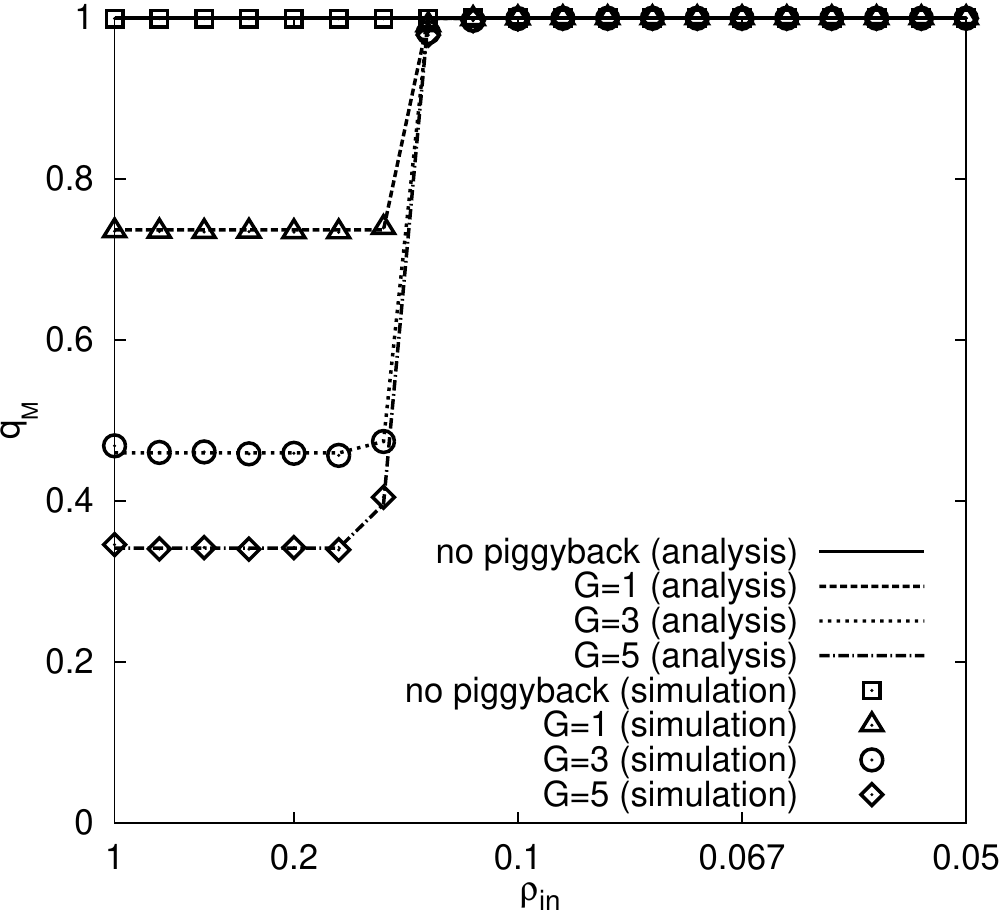}
			\label{q_vs_load_CH7}
		}		
	}
	\vspace{-12pt}
	\caption{Throughput related metrics vs $\rho_{in}$ for $L=21$ ((a)-(c)) and $L=7$ ((d)-(f)): (a),(d) Total Throughput ($\mathrm{Th}$) (b),(e) $P_{\mathrm{S}}$, and (c),(f) $q$}
	\label{Per_vs_Load_1}
	\vspace{-6pt}
\end{figure*}
\begin{figure*}[!tbp]
	{\centering
		\captionsetup[subfloat]{farskip=2pt,captionskip=1pt}
		\subfloat[Case I][]
		{
			\includegraphics[width=0.325\textwidth]{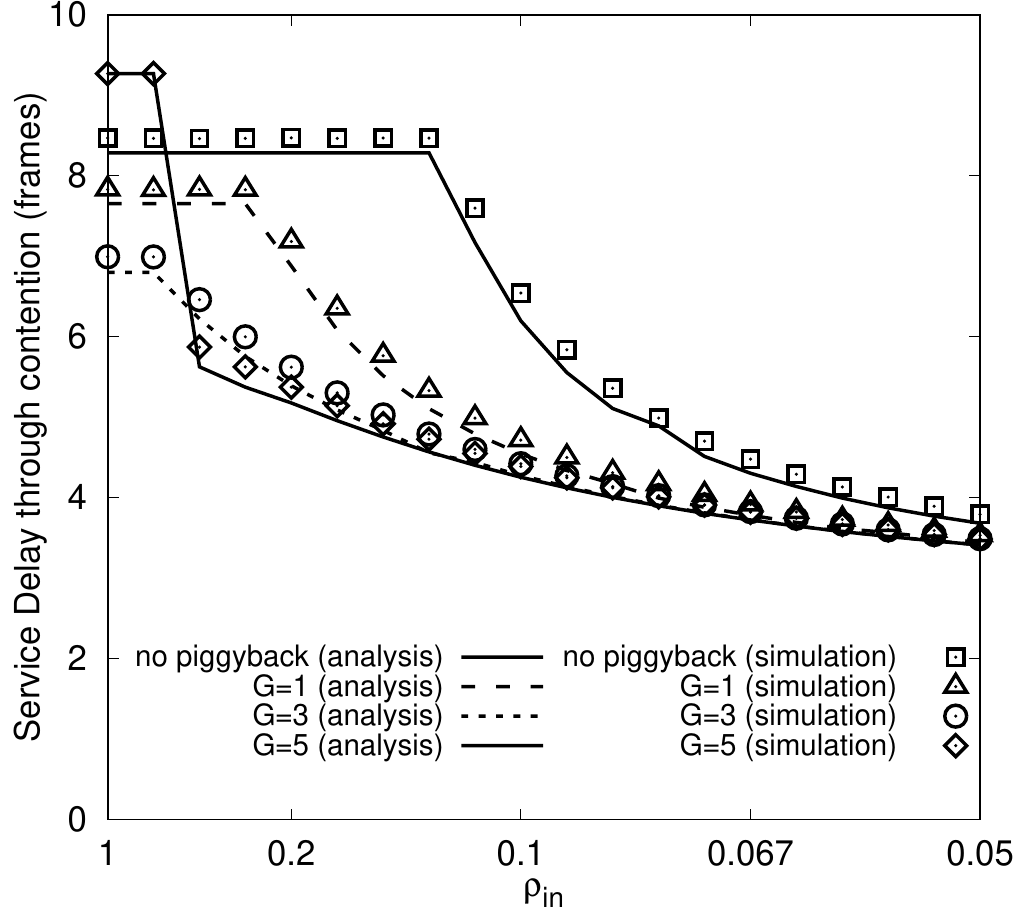}
			\label{Access_vs_load_CH21}
		}
		\subfloat[Case II][]
		{
			\includegraphics[width=0.325\textwidth]{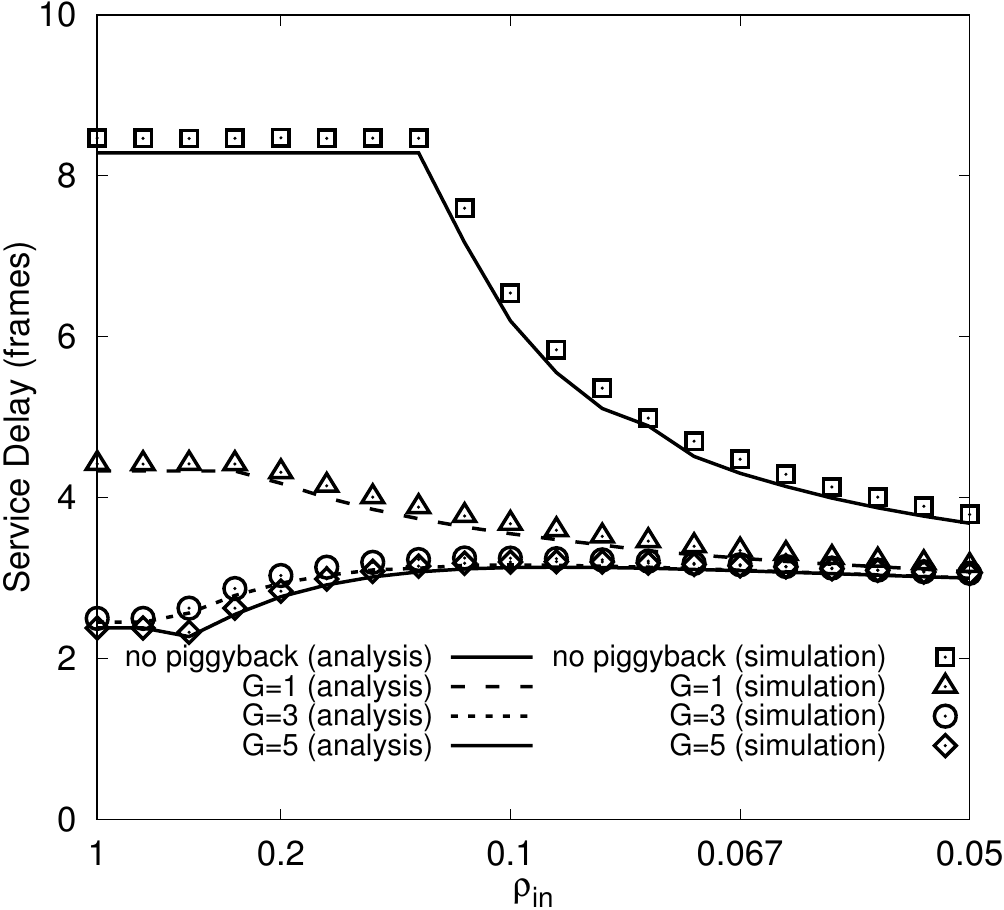}
			\label{ET_vs_load_CH21}
		}
		\subfloat[Case III][]
		{
			\includegraphics[width=0.325\textwidth]{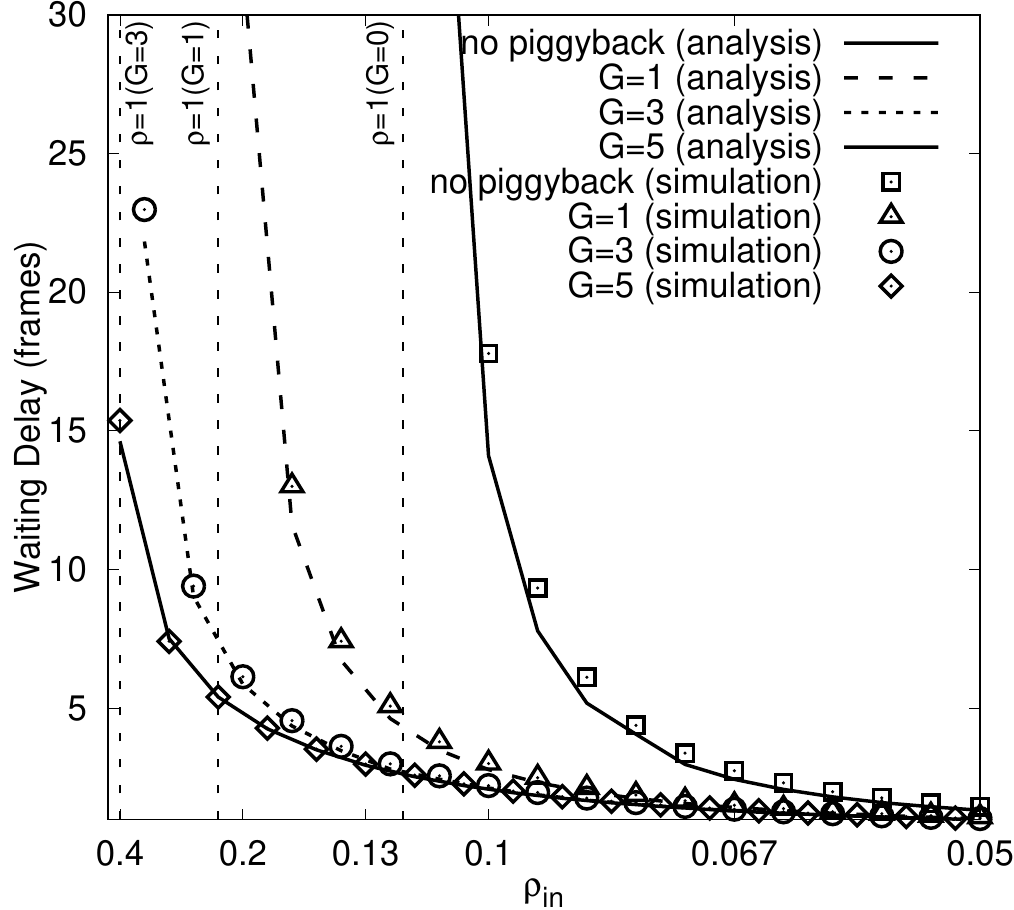}
			\label{QDelay_vs_load_CH21}
		}	
		\\
		\subfloat[Case IV][]
		{
			\includegraphics[width=0.325\textwidth]{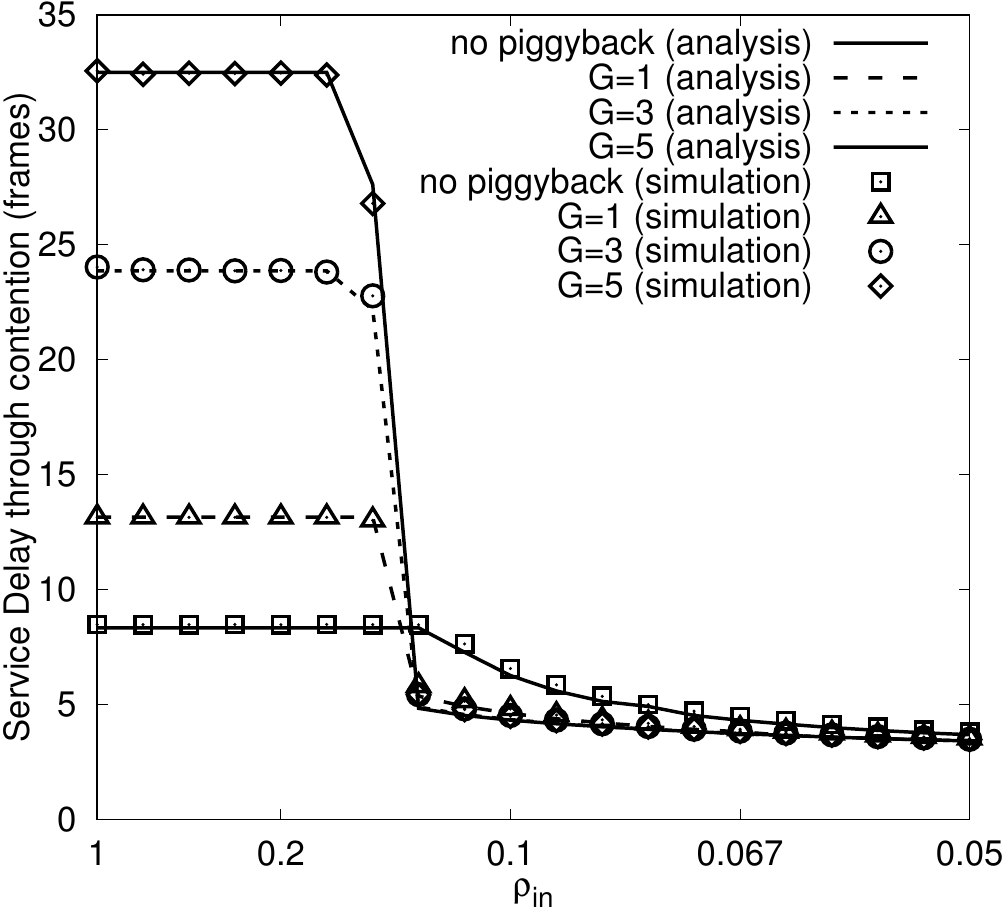}
			\label{Access_vs_load_CH7}
		}
		\subfloat[Case V][]
		{
			\includegraphics[width=0.325\textwidth]{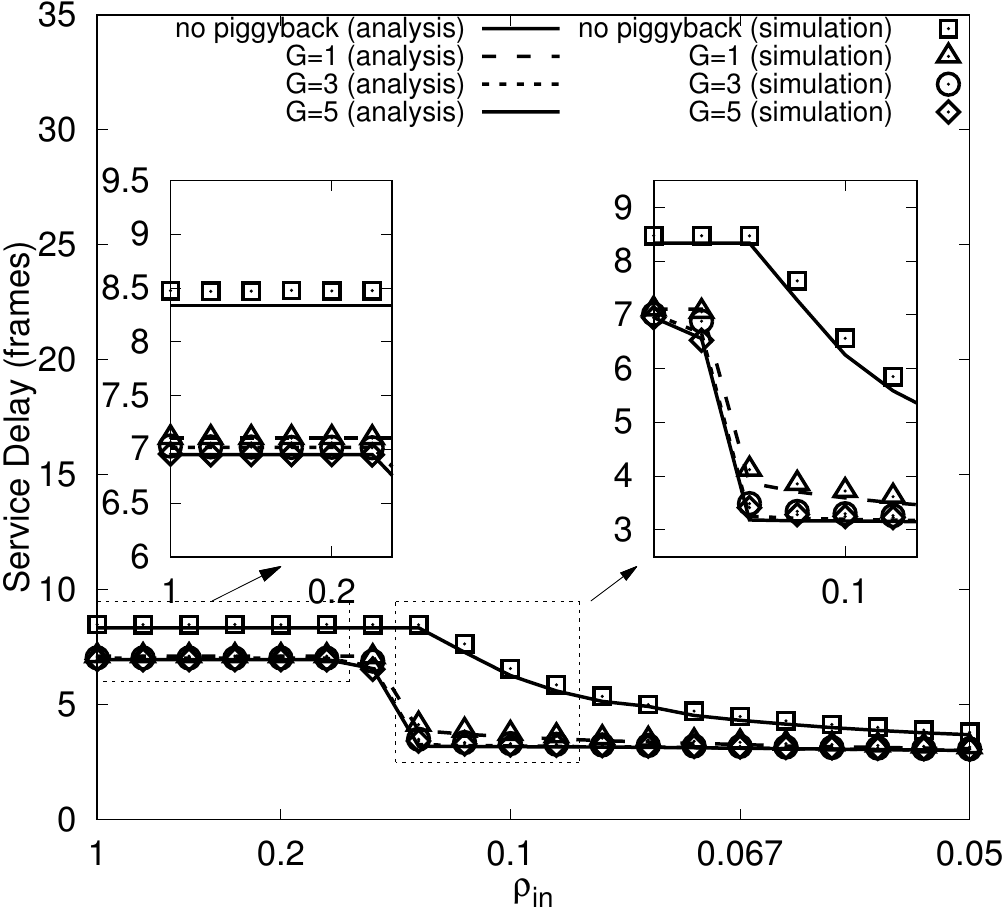}
			\label{ET_vs_load_CH7}
		}
		\subfloat[Case VI][]
		{
			\includegraphics[width=0.325\textwidth]{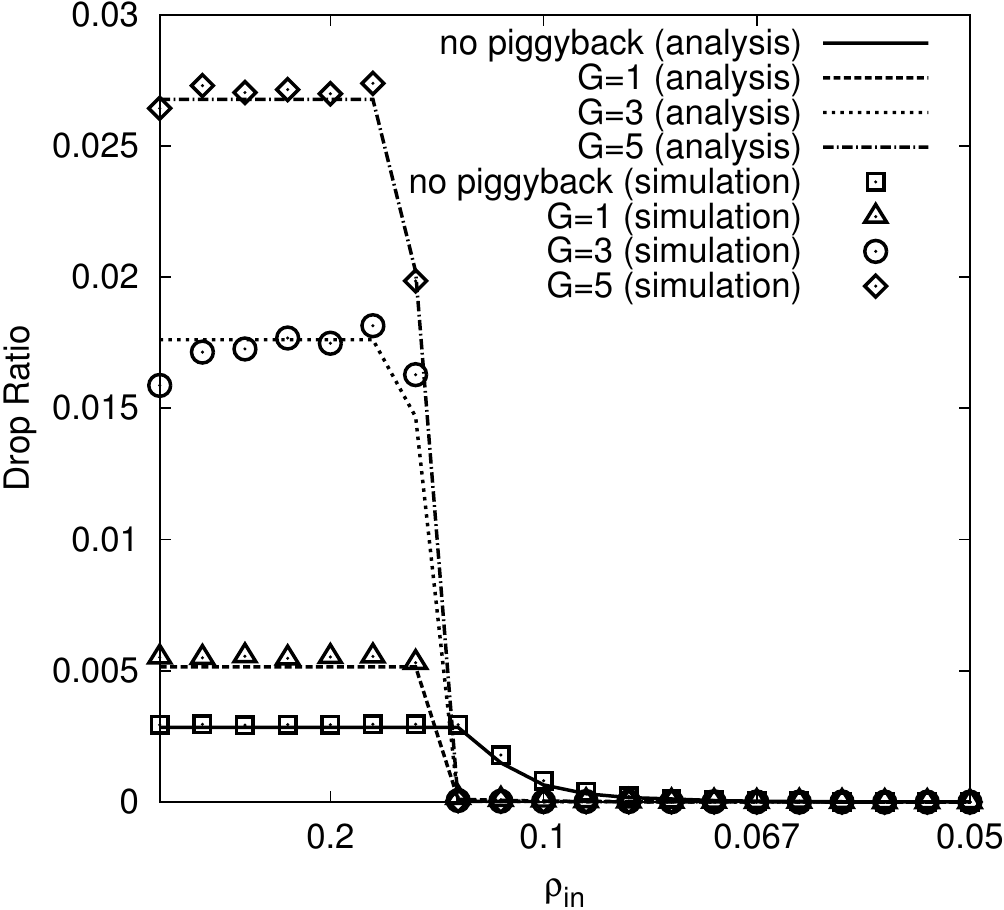}
			\label{drop_vs_load_CH7}
		}		
	}
	\vspace{-12pt}
	\caption{Delay related metrics vs $\rho_{in}$ for $L\!=\!21$ ((a)-(c)) and $L\!=\!7$ ((d)-(f)): (a),(d) Service Delay through contention ($\mathrm{E}\{S_{\mathrm{C}}\}$) (b),(e) Service Delay ($\mathrm{E}\{S\}$), (c) Waiting Delay ($\mathrm{E}\{W\}$), (f) Drop Ratio ($P_{\mathrm{D}}$)}
	\label{Per_vs_Load_2}
	\vspace{-12pt}
\end{figure*}

In the next experiment we examine the system performance with respect to the offered load. For this, we use two cases with different bandwidth availability; $L\!=\!7$ and $L\!=\!21$. As discussed in the previous experiment, the first value is the minimum that allows throughput maximization when piggyback is disabled. At the same time, $L\!=\!21$ is adequate for illustrating the performance differentiation for various values of $\mathrm{G}$. Fig.~\ref{Thput_vs_load_CH21} and~\ref{Thput_vs_load_CH7} present the total throughput for $L\!=\!21$ and $L\!=\!7$ respectively. Reasonably, in both cases, when the offered load is relatively low there is no performance differentiation regardless of whether we use piggyback (with any value of $\mathrm{G}$) or not. Since the offered load is low the probability of successive arrivals that could trigger the piggyback mechanism, i.e. those with small temporal separation, is low. The benefits of piggybacking appear when the offered load increases. However, in the case of $L\!=\!7$ the improvement is limited and the same saturation throughput is achieved for all values of $\mathrm{G}$. On the contrary, in the case of $L\!=\!21$ a greater value of $\mathrm{G}$ results in a higher maximum throughput. Note that saturation, i.e. $\rho\!=\!1$, manifests itself at an increasing level of offered load for higher values of $\mathrm{G}$ (e.g. $\rho_{in}\approx0.25$ for $\mathrm{G}\!=\!1$, $\rho_{in}\approx0.5$ for $\mathrm{G}\!=\!3$). In other words, piggybacking manages to improve system stability, i.e. $\rho\!<\!1$ for a wider range of offered load. This is in contrast to the $L\!=\!7$ case where for all values of $\mathrm{G}$ saturation appears when $\rho_{in}\approx0.14$.
The basic reason for witnessing the contrasting performance features in the two cases is BW availability. More specifically, when $L\!=\!7$ saturation is clearly the result of limited BW availability and the discussion in the previous experiment can again explain this performance. In fact, all supporting evidence can be found in Fig.~\ref{Ps_vs_load_CH7} and~\ref{q_vs_load_CH7}; $q_{_{\mathrm{M}}}$ decreases with $\mathrm{G}$ while $P_{\mathrm{S}}$ is smaller compared to the case with more available bandwidth (i.e. $L\!=\!21$, Fig.~\ref{Ps_vs_load_CH21}). On the other hand, the $L\!=\!21$ case portrays a more ``healthy'' operation mode. System saturation comes as the result of the limited capabilities of the IEEE 802.16 contention access scheme. Indeed, with the exception of $\mathrm{G}\!=\!5$ (our simulation revealed that $L\!\approx\!35$ is required to achieve maximum throughput in this case), the system allocates BW to all successfully contending BRs (Fig.~\ref{q_vs_load_CH21}) and the maximum throughput is determined by $P_{\mathrm{S}}$ which is a characteristic of the contention mechanism.

The smoother operation in the $L\!=\!21$ case is also evident in the delay-related performance metrics (Fig.~\ref{Access_vs_load_CH21}-\ref{QDelay_vs_load_CH21}). Piggybacking not only manages to reduce the overall service delay but it also significantly reduces the delay for packets transmitted through contention. Since a significant amount of traffic is forwarded using piggybacked BRs, congestion in the basic random access mechanism is alleviated. 
Furthermore, every contending BR receives a BW grant in the next frame ($q_{_{\mathrm{M}}}\!=\!1$). The immediate consequence is that an SS needs to wait for fewer frames, thus the reduced $\mathrm{E}\{S_{C}\}$. Offloading traffic through piggyback has also a positive influence on the waiting delay in the queue. Using higher $\mathrm{G}$ values drastically reduces it (Fig.~\ref{QDelay_vs_load_CH21}). Note again that the system stability is improved. i.e. $\rho\!<\!1$ for a wider range of the offered load. Concluding, when sufficient bandwidth exists it is always beneficial to use piggyback and the benefits are more important for higher $\mathrm{G}$ values. This is not the case however when the bandwidth is rather limited (e.g. $L\!=\!7$). Although increasing $\mathrm{G}$ reduces the overall delay (Fig.~\ref{ET_vs_load_CH7}) the gains are limited. More importantly, the service delay provided by the contention method increases abruptly when the offered load is high. This results in a significant delay jitter compared to the piggyback mechanism. Finally, another downside is that there is a non-negligible probability of dropping a BR (Fig.~\ref{drop_vs_load_CH7}).

So far, the discussion in the two experiments, besides the advantages of piggybacking, highlights the need for a careful consideration when choosing $\mathrm{G}$. The optimal value should always be chosen after considering the available bandwidth. Choosing a high value for $\mathrm{G}$ may have a significant impact on the smooth operation of the basic contention mechanism.

\begin{figure*}[!t]
	{\centering
		\subfloat[Case I][]
		{
			\includegraphics[width=0.33\textwidth]{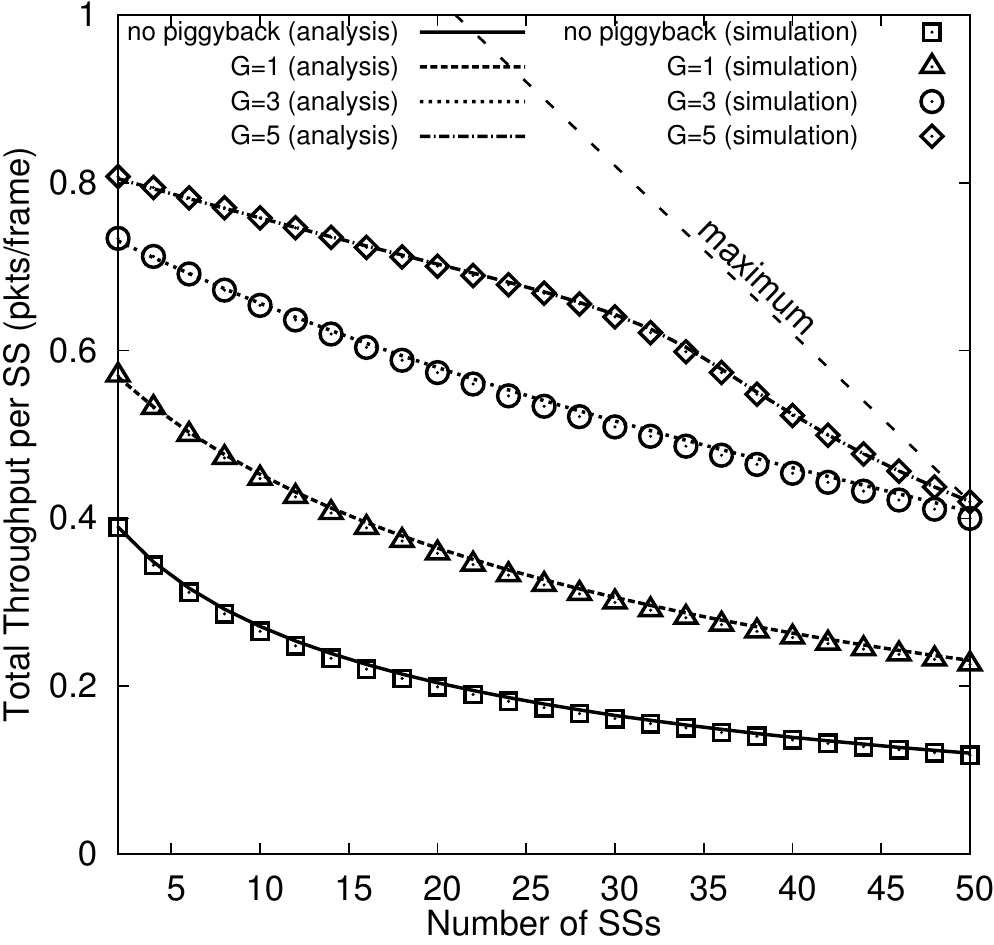}
			\label{Thput_vs_SSs_CH21}
		}
		\subfloat[Case II][]
		{
			\includegraphics[width=0.33\textwidth]{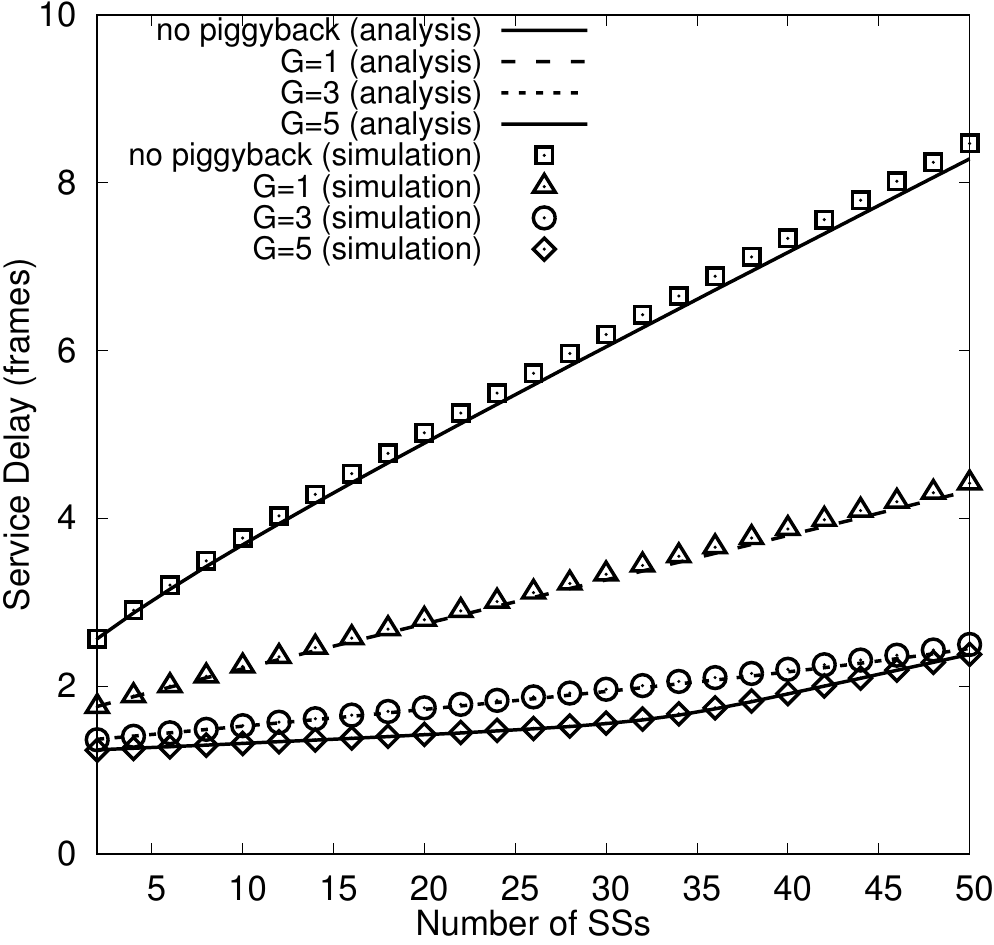}
			\label{ET_vs_SSs_CH21}
		}
		\subfloat[Case III][]
		{
			\includegraphics[width=0.33\textwidth]{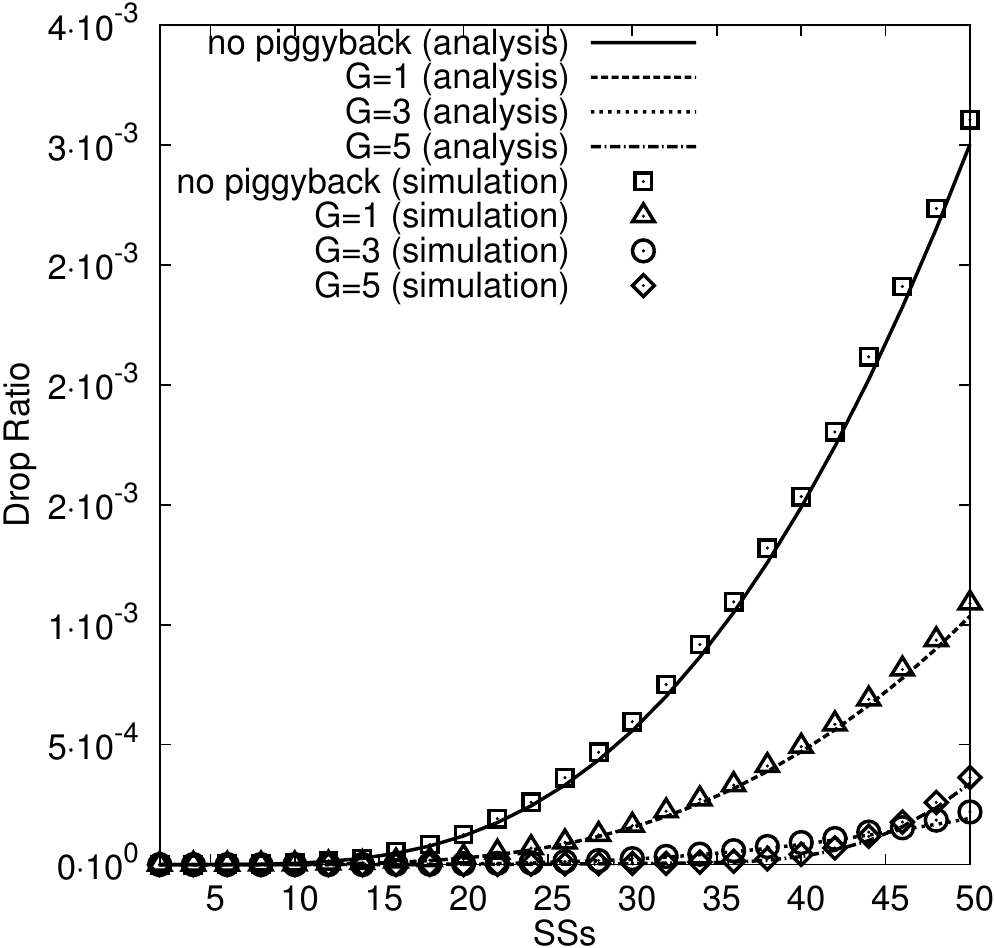}
			\label{drop_vs_SSs_CH21}
		}
	}
	\vspace{-12pt}
	\caption{Performance vs number of SSs ($N$) for $L\!=\!21$ under saturated conditions ($\rho_{in}=1$): (a) Total Throughput per SS ($\mathrm{Th}/N$) (b) Service Delay ($\mathrm{E}\{S\}$), (c) Drop Ratio ($P_{\mathrm{D}}$)}
	\label{Per_vs_SSs_1}
	\vspace{-12pt}
\end{figure*}
Finally, we examine performance against the number of SSs that participate in the network when $L\!=\!21$ and $\rho_{in}\!=\!1$ (Fig.~\ref{Per_vs_SSs_1}). Again, it is evident that the piggyback mechanism can provide a more efficient operation. In particular, although the throughput seen by each SS (Fig.~\ref{Thput_vs_SSs_CH21}) reasonably decreases as more SSs participate in the network, the use of piggyback always allows an SS to sustain a higher throughput. Similar gains are witnessed with respect to service delay (Fig.~\ref{ET_vs_SSs_CH21}) or the probability of dropping a BR (Fig.~\ref{drop_vs_SSs_CH21}). Note that the gains obtained by using $\mathrm{G}\!=\!5$ instead of $\mathrm{G}\!=\!3$ are limited when the number of SSs is close to 50. This is because in this case the system's bandwidth is not adequate for supporting such an extensive use of piggybacking (as mentioned previously, when $\mathrm{G}\!=\!5$ performance is maximized for $L\!\approx\!35$). This is another confirmation that one should consider the available BW when choosing the optimal $\mathrm{G}$.

\section{Conclusion}\label{conclusions}
In this work, we proposed and validated through extensive simulations an analysis of broadcast polling when the piggyback mechanism is employed by SSs. Unlike other analytical efforts, our model covers not only the contention phase but also the bandwidth allocation one. Moreover, it places emphasis on using a more accurate modeling of the system in general and especially for those of its aspects that are directly associated with piggybacking. This, on one hand, enables us to analyze piggybacking and its synergy with broadcast polling under more realistic conditions. On the other hand, it also facilitates a more accurate model for plain broadcast polling. The detailed performance evaluation of piggybacking reveals that it can potentially bring significant performance improvements. However, it also brings to light the associated trade-offs regarding the delay jitter as well as the need for a careful allocation of UL BW between contending and piggybacked BRs.

\appendices
\setcounter{equation}{0}
\renewcommand{\theequation}{\thesection.\arabic{equation}}

\section{Probability of an Empty Queue seen by a Departing customer}\label{appendix-mg1}
In a generic M/G/1 system with vacations the probability that an arbitrary departing customer leaves the system empty is given by\cite{kleinrock1975queueingsystems}:
\vspace{-6pt}
\begin{equation}\label{klein-basic}
\Pi_{0}=(1-\rho)/F\acute{}(1)
\end{equation}
where $F(z)=\sum_{j=1}^{\infty}f_{j}z^{j}$ is the probability generating function for the number of customers awaiting service when the server returns from vacation to find at least one customer waiting. In other words,
\begin{equation}
\begin{split}
f_{j} &= \mathrm{P}\{j \text{ customers waiting after vacation } | j\neq 0\} \\
&=\dfrac{\mathrm{P}\{j \text{ customers arrive during last vacation} \}}{\mathrm{P}\{j\neq 0\}} \nonumber
\end{split}
\end{equation}
Note that the number of customers waiting at the end of vacation is the number of customers that arrived during the last vacation interval because at the beginning of that last interval the number of waiting customers was zero (otherwise there would be no vacation). In our case the vacation duration is deterministic with value $\mathrm{T_{fr}}$. Therefore, 
\begin{equation}
f_{j} = \dfrac{\mathrm{P}(j;\lambda\mathrm{T_{fr}})}{1-e^{\lambda\mathrm{T_{fr}}}} \nonumber
\end{equation}
where $\mathrm{P}(j;\lambda\mathrm{T_{fr}})$ is the probability of $j$ arrivals during a frame. Note that $F\acute{}(1)\!=\!\mathrm{E}(j)$ and
\begin{equation}
\mathrm{E}(j) = \dfrac{\sum_{j=1}^{\infty} j\mathrm{P}(j;\lambda\mathrm{T_{fr}}) }{1-e^{\lambda\mathrm{T_{fr}}}} \nonumber
\end{equation}
Note that the sum in the previous equation is equal to the mean of Poisson distribution, therefore
\begin{equation}
F\acute{}(1) = \dfrac{\lambda\mathrm{T_{fr}}}{1-e^{-\lambda\mathrm{T_{fr}}}}
\end{equation}
which in combination with (\ref{klein-basic}) results in (\ref{empty_queue}).

\setcounter{equation}{0}
\section{Solving the Markov Chain}\label{appendix-markov}
\begin{figure*}[!b]
	\hrulefill
	\newcounter{MYtempeqncnt2}
	\setcounter{MYtempeqncnt2}{\value{equation}}
	\setcounter{equation}{10}
	\begin{equation}\label{pidle}
		\begin{split}
			\mathrm{P_{Idle}} & =
			p_{a}\mathrm{P_{Idle}}\!+\!\Pi_{0}\sum_{i=0}^{\mathrm{D}}\sum_{j=1}^{\mathrm{M}}b^{\mathrm{T}}_{i,j}\!+\!\Pi_{0}\sum_{k=1}^{\mathrm{G}}\mathrm{P}\{\mathrm{PG}_{i}\}\!+\!\Pi_{0}(b^{\mathrm{C}}_{D,M}\!+\!b^{\mathrm{F}}_{D,M})
			\stackrel{\substack{(\ref{basic_pf}),(\ref{success_states}),(\ref{piggy_states})}}{\xlongequal{\;\;}} p_{a}\mathrm{P_{Idle}}\!+\!\Pi_{0}(1\!-\!p_{f})\tau b_{0,0}^{\mathrm{R}}\\
			+ & [1\!-\!(1\!-\!\Pi_{0})^{\mathrm{G}}](1\!-\!\Pi_{0})(1\!-\!p_{f})\tau b_{0,0}^{\mathrm{R}}\!+\!\Pi_{0}(p_{f})^{\mathrm{D}\!+\!1}b_{0,0}^{\mathrm{R}} =\overbrace{\frac{\Pi_{0}+[1\!-\!(1\!-\!\Pi_{0})^{\mathrm{G}}](1-\Pi_{0})[1\!-\!(p_{f})^{\mathrm{D}+1}]}{1\!-\!p_{a}}}^{\Phi}b_{0,0}^{\mathrm{R}}
		\end{split}
	\end{equation}
	\begin{equation}\label{normalization}
		\begin{split}
			1 = &\sum_{i=1}^{\mathrm{D}}\sum_{j=-\overline{\mathrm{K}}_{i}}^{-1}b_{i,j}^{\mathrm{W}}+\sum_{i=0}^{\mathrm{D}} b_{i,0}^{\mathrm{R}}+\sum_{i=0}^{\mathrm{D}}\sum_{j=1}^{\mathrm{M}}b_{i,j}^{\mathrm{C}}+\sum_{i=0}^{\mathrm{D}}\sum_{j=1}^{\mathrm{M}}b_{i,j}^{\mathrm{T}}+\sum_{i=0}^{\mathrm{D}}\sum_{j=1}^{\mathrm{M}}b_{i,j}^{\mathrm{F}}+\sum_{i=1}^{\mathrm{G}}\mathrm{\mathrm{P}}\{\mathrm{PG}_{i}\}+\mathrm{P_{Idle}}\\
			\stackrel{\substack{(\ref{basic_pf})-(\ref{pidle})}}{\xlongequal{\;\;}}& \bigg\{\tau\!+\!\Omega\!+\!p\mathrm{M}\tau\!+\!(1\!-\!p_{f})\tau\!+\!\frac{1\!-\!q}{q}(1\!-\!p_{f})\tau\!+\!Z\!+\!\Phi\bigg\}b_{0,0}^{\mathrm{R}}\Rightarrow
			b_{0,0}^{\mathrm{R}}\!=\!\frac{1}{\tau\big(1\!+\!p\mathrm{M}\!+\!\frac{1\!-\!p_{f}}{q}\big)\!+\!\Omega\!+\!Z\!+\!\Phi}
		\end{split}
	\end{equation}
	\setcounter{equation}{\value{MYtempeqncnt2}}
\end{figure*}

Regarding the states corresponding to the transmission of a BR, i.e. $(i,0)^{\mathrm{R}}, \forall i \in [0,D]$, recall that the SS moves to contention round $i+1$ and transmits again a BR if it fails to transmit data in round $i$. This happens because either the SS did not receive a BW grant in $\mathrm{M}$ consecutive frames or its BR is caught up in a collision. Therefore the corresponding probability is $p_{f}=p+(1-p)(1-q)^{\mathrm{M}}$. Indeed, using the proposed Markov chain (Fig.~\ref{markov_chain}) we can show that $b^{R}_{i+1,0}=p_{f}b^{R}_{i,0}$. By recursion we establish that:
\begin{equation}\label{basic_pf}
b^{R}_{i,0}=p^{i}_{f}b^{R}_{0,0}, \forall i\in [1,\mathrm{D}]
\end{equation}
and therefore:
\vspace{-6pt}
\begin{equation}\label{trans_slot}
\sum_{i=0}^{D}b_{i,0}^{\mathrm{R}}=\sum_{i=0}^{\mathrm{D}}(p_{f})^{i}b_{0,0}^{\mathrm{R}}=\overbrace{\frac{1-(p_{f})^{\mathrm{D}+1}}{1-p_{f}}}^{\tau}b_{0,0}^{\mathrm{R}}=\tau b_{0,0}^{\mathrm{R}}
\end{equation}
For the states traversed during back-off $b^{\mathrm{W}}_{i,j}=b^{\mathrm{R}}_{i,0}, \forall j\in [-\overline{\mathrm{K}}_{i},-1]$. Using (\ref{basic_pf}):
\vspace{-6pt}
\begin{equation}
\sum_{i=0}^{\mathrm{D}}\sum_{j=-\overline{\mathrm{K}}_{i}}^{-1}\!\!b_{i,j}^{\mathrm{W}}\!=\!\sum_{i=1}^{\mathrm{D}}\overline{\mathrm{K}}_{i}b_{i,0}^{\mathrm{R}}=\overbrace{\sum_{i=1}^{\mathrm{D}}(p_{f})^{i}\overline{\mathrm{K}}_{i}}^{\Omega}b_{0,0}^{\mathrm{R}}=\Omega b_{0,0}^{\mathrm{R}}
\end{equation}
In the case of a collision in round $i \in [0,\mathrm{D}]$ the SS will go over states $(i,j)^{\mathrm{C}}, j\in [1,\mathrm{M}]$. Note that $b^{\mathrm{C}}_{i,j}=b^{\mathrm{C}}_{i,k}, \forall j,k \in [1,\mathrm{M}]$. The probability of collision is $p$ therefore $b^{\mathrm{C}}_{i,1}=pb^{\mathrm{R}}_{i,0}$. With the help of (\ref{basic_pf}) we find that:
\begin{equation}
\sum_{i=0}^{\mathrm{D}}\sum_{j=1}^{\mathrm{M}}b_{i,j}^{\mathrm{C}}=p\mathrm{M}\sum_{i=0}^{\mathrm{D}}(p_{f})^{i}b_{0,0}^{\mathrm{R}}=p\mathrm{M}\tau b_{0,0}^{\mathrm{R}}
\end{equation}
In the case of a successful BR the possible states are $(i,j)^{\mathrm{T}}$ and $(i,j)^{\mathrm{F}}, \forall j\in [1,\mathrm{M}], i\in [0,\mathrm{D}]$ and the corresponding probabilities result from the Markov chain as:
\begin{align}\label{success_and_fail}
b^{\mathrm{T}}_{i,j} &= (1\!-\!p)q(1\!-\!q)^{j\!-\!1}b_{i,0}^{\mathrm{R}}, & \forall j\in [1,\mathrm{M}], i\in [0,\mathrm{D}]\\
b^{\mathrm{F}}_{i,j} &= (1\!-\!p)(1\!-\!q)^{j}b_{i,0}^{\mathrm{R}}, & \forall j\in [1,\mathrm{M}], i\in [0,\mathrm{D}] \nonumber
\end{align}
Combining these equations with (\ref{basic_pf}), we show that:
\begin{equation}\label{success_states}
\begin{split}
\sum_{i=0}^{\mathrm{D}}\sum_{j=1}^{\mathrm{M}}b_{i,j}^{\mathrm{T}} & =\sum_{i=0}^{\mathrm{D}}\sum_{j=1}^{\mathrm{M}}q(1\!-\!p)(1\!-\!q)^{j-1}p^{i}_{f}b_{0,0}^{\mathrm{R}}\\
& =(1\!-\!p_{f})\tau b_{0,0}^{\mathrm{R}}
\end{split}
\end{equation}
\begin{equation}\label{fail_states}
\begin{split}
\sum_{i=0}^{\mathrm{D}}\sum_{j=1}^{\mathrm{M}}b_{i,j}^{\mathrm{F}} & =\sum_{i=0}^{\mathrm{D}}\sum_{j=1}^{\mathrm{M}}(1\!-\!p)(1\!-\!q)^{j}p^{i}_{f}b_{0,0}^{\mathrm{R}}\\
& =\frac{1\!-\!q}{q}(1\!-\!p_{f})\tau b_{0,0}^{\mathrm{R}}
\end{split}
\end{equation}
Regarding the piggyback states, note that:
\begin{equation}\label{first_piggy_state}
\begin{split}
&\mathrm{\mathrm{P}}\{\mathrm{PG}_{1}\}=\sum_{i=0}^{\mathrm{D}}\sum_{j=1}^{\mathrm{M}}(1\!-\!\Pi_{0})b_{i,j}^{\mathrm{T}}\\
& = (1\!-\!\Pi_{0})(1\!-\!p_{f})\tau b_{0,0}^{\mathrm{R}} 
\end{split}
\end{equation}
\begin{equation}
		\mathrm{\mathrm{P}}\{\mathrm{PG}_{i}\}=(1\!-\!\Pi_{0})\mathrm{\mathrm{P}}\{\mathrm{PG}_{i\!-\!1}\}, \forall i \in [2,\mathrm{G}]
\end{equation}
therefore:
\begin{equation}\label{piggy_states}
\begin{split}
\sum_{i=1}^{\mathrm{G}}\mathrm{\mathrm{P}}\{\mathrm{PG}_{i}\} & = \mathrm{\mathrm{P}}\{\mathrm{PG}_{1}\}\sum_{i=1}^{\mathrm{G}}(1\!-\!\Pi_{0})^{i-1}\\
& =\overbrace{(1\!-\!\Pi_{0})(1\!-\!p_{f})\tau\frac{1-(1\!-\!\Pi_{0})^{\mathrm{G}}}{\Pi_{0}}}^{Z}b_{0,0}^{\mathrm{R}} 
\end{split}
\end{equation}
The probability of being in the idle state can be calculated according to (\ref{pidle}). Finally, using the normalization condition for the Markov chain, we compute $b_{0,0}^{\mathrm{R}}$ in (\ref{normalization}).

\setcounter{equation}{0}

\section{Derivation of $\mathrm{E}\{\mathcal{P}\}$ and $\mathrm{E}\{\mathcal{R}\}$}\label{appendix-q-related}

Let us first focus on $\mathrm{E}\{\mathcal{R}\}$ and assume that we examine the most recent frame (frame 0) from a sequence of $\mathrm{M}$ frames with indexes from $-(\mathrm{M}\!\!-\!\!1)$ to 0. Furthermore, let $\mathcal{R}_{-i}$ denote the number of new successful BRs through contention in frame $-i$ and $\mathcal{R}\acute{}_{-i}$ the subset of those BRs that are not served until frame 0 although waiting for $i$ frames. Then, $\mathcal{R}$ can be expressed as the sum of new successful BRs in frame 0 as well as BRs from previous frames that are yet to be served although waiting for up to $\mathrm{M}-1$ frames, i.e., $\mathcal{R}=\mathcal{R}_{0}+\mathcal{R}\acute{}_{-1}+\ldots+\mathcal{R}\acute{}_{-(\mathrm{M}-1)}$. Note that $\mathcal{R}_{0}$ is equivalent to the number of TOs in frame 0 that contain a successful BR. Since SSs act independently and choose a TO randomly, $\mathcal{R}_{0}$ can be modelled as a binomial RV, i.e., $\mathcal{R}_{0}\sim B(N_{s}, \mathrm{P}_{S})$, where $B(\cdot)$ denotes the binomial distribution, $N_{s}$ is the number of TOs and $\mathrm{P}_{S}$ is the probability of a successful BR in a TO and given in (\ref{ps_value}). Observe that, in analogy, every $\mathcal{R}_{-i}$ is also binomially distributed with the same parameters. 
Regarding $\mathcal{R}\acute{}_{-1}$, recall that in frame -1 there are $\mathcal{R}_{-1}$ new successful BRs. Each of those BRs will not be served with probability $1\!-\!q$ and will now wait for BW allocation in frame 0. The decision for serving or not one of the $\mathcal{R}_{-1}$ BRs can be modelled with a Bernoulli RV. Since the decisions for all BRs are independent, the number of non-served BRs $\mathcal{R}\acute{}_{-1}$ is a binomial RV with parameters $\mathcal{R}_{-1}$ and $1-q$, i.e., $\mathcal{R}\acute{}_{-1}\!\sim\!B(\mathcal{R}_{-1}, 1\!-\!q)$. Since $\mathcal{R}_{-1}$ is also a binomial RV with parameters $N_{s}$ and $\mathrm{P}_{S}$ it is well known that $\mathcal{R}\acute{}_{-1}\!\sim\!B(N_{s}, P_{\mathrm{S}}(1\!-\!q))$, i.e., $\mathcal{R}\acute{}_{-1}$ follows a binomial distribution with parameters $N_{s}$ and $P_{\mathrm{S}}(1-q)$. Following a similar reasoning we conclude that:
\begin{equation}
	\mathcal{R}\acute{}_{-i}\!\sim\!B(N_{s}, P_{\mathrm{S}}(1\!-\!q)^{i}), \forall i \in [1, \mathrm{M}\!-\!1] \nonumber
\end{equation}
because successful BRs in frame $-i$ reach frame 0 if and only if they do not receive a BW grant for $i$ consecutive frames. As a result, the expected number of BRs awaiting BW allocation is 
\begin{equation}\label{mean-r-calligr}
	\begin{split}
		\mathrm{E}\{\mathcal{R}\}&=\mathrm{E}\{\mathcal{R}_{0}\}+\mathrm{E}\{\mathcal{R}\acute{}_{-1}\}+\cdots+\mathrm{E}\{\mathcal{R}\acute{}_{-(\mathrm{M}\!-\!1)}\}\\
		&=N_{s}P_{\mathrm{S}}\!+\!N_{s}P_{\mathrm{S}}(1\!-\!q)\!+\!\cdots\!+\!N_{s}P_{\mathrm{S}}(1\!-\!q)^{\mathrm{M}\!-\!1}
	\end{split}	
\end{equation}
which results in (\ref{er-formula}).

Following a similar approach, we can express $\mathcal{P}$, i.e., the number of data slots allocated to piggypacked BRs, as $\mathcal{P}=\mathcal{P}_{0}+\mathcal{P}\acute{}_{-1}+\ldots+\mathcal{P}\acute{}_{-(\mathrm{G}-1)}$, where $\mathcal{P}_{-i}, \; i \in [0, \mathrm{G}-1]$ is the number of SSs that start piggyback in frame $-i$ and $\mathcal{P}\acute{}_{-i}$ is the number of SSs that start piggyback in frame $-i$ but are still in piggyback mode during frame 0. Observe that $\mathcal{P}_{0}$ is the number of SSs that had a pending BR (i.e. a successfully contending BR that awaits BW allocation) in frame -1, received BW allocation (probability $q$) and after completing the data transmission there was at least one packet in their queue (probability $1-\Pi_{0}$). Since the number of pending BRs in a frame is $\mathcal{R}$ it is clear that $\mathcal{P}_{0}\sim B(\mathcal{R},q(1-\Pi_{0}))$. Therefore $\mathrm{E}[\mathcal{P}_{0}]=\mathrm{E}[\mathcal{R}]q(1-\Pi_{0})$. 
Similarly, $\mathcal{P}_{-1}$ is the number of SSs that start piggyback in frame -1, therefore $\mathcal{P}_{-1}\sim B(\mathcal{R},q(1-\Pi_{0}))$. Only a subset of these SSs will still be in piggyback in frame 0, i.e. those that find a new queued packet after transmitting the previous one. Therefore, $\mathcal{P}\acute{}_{-1}\sim B(\mathcal{P}_{-1},1-\Pi_{0})$ or equivalently  $\mathcal{P}\acute{}_{-1}\sim B(\mathcal{R},q(1\!-\!\Pi_{0})^{2})$. As a result, $\mathrm{E}[\mathcal{P}\acute{}_{-1}]=\mathrm{E}[\mathcal{R}]q(1-\Pi_{0})^{2}$. By generalizing $\mathrm{E}[\mathcal{P}\acute{}_{-i}]=\mathrm{E}[\mathcal{R}]q(1-\Pi_{0})^{i+1}$, therefore 
\begin{equation}\label{mean-p-calligr}
\begin{split}
\mathrm{E}\{\mathcal{P}\}&=\mathrm{E}\{\mathcal{P}_{0}\}+\mathrm{E}\{\mathcal{P}\acute{}_{-1}\}+\cdots+\mathrm{E}\{\mathcal{P}\acute{}_{-(\mathrm{G}\!-\!1)}\}\\
&=\mathrm{E}\{\mathcal{R}\}q\big[(1\!\!-\!\Pi_{0})\!+\!(1\!\!-\!\Pi_{0})^{2}\!\!+\!\cdots\!+\!\!(1\!\!-\!\Pi_{0})^{\mathrm{G}}\big]
\end{split}	
\end{equation}
which results in (\ref{ep-formula}).

\setcounter{equation}{0}

\section{Derivation of Expected Waiting Delay}\label{appendix-waiting-delay}

To prove the approximation formula in (\ref{waiting-delay}) let us start with the mean value analysis used to prove the Pollaczek-Khinchin formula~\cite{cooper1981queueingtheory}. In an M/G/1 system without vacations:
\begin{equation}\label{eq:mean-value}
\begin{split}
\mathrm{E}\{W_{\mathrm{noV}}\}=\rho\mathrm{E}\{S_{\mathrm{R}}\}\!+\!N_{Q}\mathrm{E}\{S_{\mathrm{Q}}\}
=\frac{\lambda\mathrm{E}\{S^{2}\}}{2(1\!-\!\lambda\mathrm{E}\{S_{\mathrm{Q}}\})}
\end{split}	
\end{equation}
where $\mathrm{E}\{S_{\mathrm{R}}\}$ is the remaining service time of the customer in service seen by an arriving customer, $N_{Q}$ is the expected number of customers in queue and $\mathrm{E}\{S_{\mathrm{Q}}\}$ the expected service delay for customers in the queue. We have also used $N_{Q}=\lambda\mathrm{E}\{W_{\mathrm{noV}}\}$ from Little's law and $\mathrm{E}\{S_{\mathrm{R}}\}\!=\!\lambda\mathrm{E}\{S^{2}\}/2\mathrm{E}\{S\}$~\cite{cooper1981queueingtheory}. In a typical M/G/1 system $\mathrm{E}\{S_{\mathrm{Q}}\}\!=\!\mathrm{E}\{S\}$ and the Pollaczek-Khinchin formula follows from (\ref{eq:mean-value}). When server vacations of fixed size $\mathrm{T_{fr}}$ are considered it can be proved that $\mathrm{E}\{W\}\!=\!\mathrm{E}\{W_{\mathrm{noV}}\}\!+\!\mathrm{T_{fr}}/2$~\cite{cooper1981queueingtheory} and in combination with (\ref{eq:mean-value}) we receive (\ref{waiting-delay}) where $\mathrm{E}\{S^{2}\}$ can be approximated using (\ref{mean_service_delay}) as
\begin{equation}\label{mean_square_service_delay}
\begin{split}
\mathrm{E}\{S^{2}\}\!\!\approx\!\!\dfrac{(p_{f})^{\mathrm{D}\!+\!1}\!(\mathrm{E}\{S_{\mathrm{D}}\})^{2}\!\!+\!\!\sum\limits_{i=0}^{\mathrm{D}}\!\sum\limits_{j=1}^{\mathrm{M}} \!bb_{i,j} (\mathrm{E}\{S_{i,j}\})^{2}\!\!+\!\!Z}{1+Z}
\end{split}
\end{equation}

Unlike a typical M/G/1 system, in our case the expected service delay for customers in the queue is not the same as the overall expected service delay, i.e., $\mathrm{E}\{S_{\mathrm{Q}}\}\neq\mathrm{E}\{S\}$. This is because only queued packets can enjoy the low delay of piggybacking. To determine $\mathrm{E}\{S_{\mathrm{Q}}\}$ recall that an arriving packet enters the queue with probability $1\!-\!\Pi_{0}$. In such a case, the new packet finds the system either busy (with probability $\rho$) or idle (with probability $1\!-\!\Pi_{0}\!-\!\rho$), i.e. there are other packets waiting but the server is still in vacation. Therefore 
\begin{equation}\label{eq:inqueue-delay}
	\mathrm{E}\{S_{\mathrm{Q}}\}=\frac{\rho}{1-\Pi_{0}}\mathrm{E}\{S_{B}\}+\frac{1-\Pi_{0}-\rho}{1-\Pi_{0}}\mathrm{E}\{S_{I}\}
\end{equation}
where $\mathrm{E}\{S_{B}\}$and $\mathrm{E}\{S_{I}\}$denote the service delay experienced by the arriving packet in each of the aforementioned cases. In the case that the packet finds the system non empty but idle it is clear that the packet is not the first one waiting for service. This means that with high probability it will be served through piggybacking therefore $\mathrm{E}\{S_{I}\}\simeq 1$ unless $\mathrm{G}\!=\!0$, i.e. piggyback is disabled, in which case $\mathrm{E}\{S_{I}\}\simeq \mathrm{E}\{S_{\mathrm{C}}\}$. Regarding $\mathrm{E}\{S_{B}\}$, we can rewrite the overall service delay 
\begin{equation}\label{eq:busy-determination}
	\mathrm{E}\{S\}=\Pi_{0}\mathrm{E}\{S_{\mathrm{C}}\}+(1\!-\!\Pi_{0}\!-\!\rho)\mathrm{E}\{S_{I}\}+\rho\mathrm{E}\{S_{B}\}
\end{equation}
where the first component corresponds to an arriving packet that finds the system empty (probability $\Pi_{0}$) in which case it will be served through contention. It is possible to determine $\mathrm{E}\{S_{B}\}$ using (\ref{delay_contention}) and (\ref{eq:busy-determination}).


\end{document}